\documentclass[amsmath,amssymb,aps,twocolumn]{revtex4-2}

\usepackage{graphicx}% Include figure files
\usepackage{epstopdf}
\usepackage{dcolumn}% Align table columns on decimal point
\usepackage{bm}% bold math
\usepackage{epsf}
\usepackage{epsfig}
\usepackage{verbatim}
\usepackage{amsmath}
\usepackage{amsfonts}
\usepackage{ifthen}
\usepackage[normalem]{ulem}
\usepackage[flushleft]{threeparttable}
\usepackage{aas_macros}
\usepackage[percent]{overpic}
\usepackage{tabularx}
\usepackage{xcolor}
\definecolor{darkgreen}{rgb}{0.0,0.5,0.0}
\usepackage[colorlinks,citecolor=darkgreen]{hyperref} % Works with PDFLaTeX; use for arXiv and ShareLaTeX

\newcommand{\eg}{\emph{e.g.,} }

\newcommand{\be}{\begin{equation}}
\newcommand{\ee}{\end{equation}}
\newcommand{\bea}{\begin{equation*}}
\newcommand{\eea}{\end{equation*}}
\newcommand{\beq}{\begin{equation} }
\newcommand{\eeq}{\end{equation}}
\newcommand{\beqr}{\begin{eqnarray} \nonumber}
\newcommand{\eeqr}{\end{eqnarray}}
\newcommand{\beqrb}{\begin{eqnarray}}
\newcommand{\eeqrb}{\nonumber \end{eqnarray}}

\newcommand{\coma}{\mbox{ ,}}

\newcommand{\cm}{\mbox{ cm}}
\newcommand{\sr}{\mbox{ sr}}
\newcommand{\se}{\mbox{ s}}

\newcommand{\erg}{\mbox{ erg}}

\newcommand{\MHz}{\mbox{ MHz}}
\newcommand{\GHz}{\mbox{ GHz}}

\newcommand{\Mpc}{\mbox{ Mpc}}

\newcommand{\keV}{\mbox{ keV}}

\newcommand{\GeV}{\mbox{ GeV}}

\newcommand{\mK}{\mbox{ mK}}

\newcommand{\const}{\mbox{const}}

\newcommand{\dgr}{^{\circ}}
\newcommand{\dgrdot}{{\overset{^\circ}{.}}}

\newcommand{\Myfr}{{\nu}}
\newcommand{\Myc}{{\mathsf{c}}}
\newcommand{\MyH}{{\mathsf{h}}}
\newcommand{\MyT}{{T_b}}

\newcommand{\MySig}{{S}}
\newcommand{\pxN}{{\mathsf{n}}}
\DeclareMathOperator{\arccsc}{arccsc}

\begin{document}

\title{Radially polarized synchrotron from galaxy-cluster virial shocks}

\author{Uri Keshet}
\affiliation{Physics Department, Ben-Gurion University of the Negev, POB 653, Be'er-Sheva 84105, Israel; keshet.uri@gmail.com}

\begin{abstract}
Radio-to-$\gamma$-ray signals, recently found narrowly confined near the characteristic $2.4R_{500}$ scaled radii of galaxy clusters and groups, have been associated with their virial (structure-formation accretion) shocks based on spectro-spatial characteristics.
By stacking high-latitude GMIMS radio data around MCXC galaxy clusters, we identify ($3\sigma$--$4\sigma$) excess radially polarized emission at the exact same scaled radius, providing directional support, and indicating tangential magnetic fields induced by the shocked inflow.
The results suggest a strong mass dependence, a flat energy spectrum, and a high polarization fraction, consistent with synchrotron emission from electrons accelerated by strong virial shocks.
The narrow radial range of such stacked virial-shock signals suggests that although the shocks are theorized to have diverse, irregular morphologies, they share similar $\sim2.4R_{500}$ minimal radii.
\end{abstract}

\maketitle

\section{Introduction}
\label{sec:Intro}

Searches for the faint radiative signatures predicted to arise from virial, i.e. accretion, or structure-formation, shocks \citep{LoebWaxman00, TotaniKitayama00, WaxmanLoeb00, KeshetEtAl03, Miniati02, KeshetEtAl04, KeshetEtAl04_SKA, KocsisEtAl05} turned out successful once data were stacked over many clusters and groups of galaxies (henceforth: clusters), after \citep{ReissEtAl17, reiss2018detection, HouEtAl23, Keshet25PaperII, Keshet25PaperIII} but not before (\eg \citep{HuberEtAl13, AckermannEtAl14_GammaRayLimits, ProkhorovChurazov14, GriffinEtAl14}) lengthscales were normalized by the characteristic $R_{500}$ radius of each cluster, thus utilizing the approximate self-similarity of the intracluster medium (ICM). Here, $R_{500}$ encloses a mean density $500$ times the critical mass density of the Universe.

\begin{figure}[h!]
    \vspace{-0.4cm}
    \centerline{
    \includegraphics[width=0.87\linewidth,trim={0cm 0.03cm 0cm 0.02cm},clip]{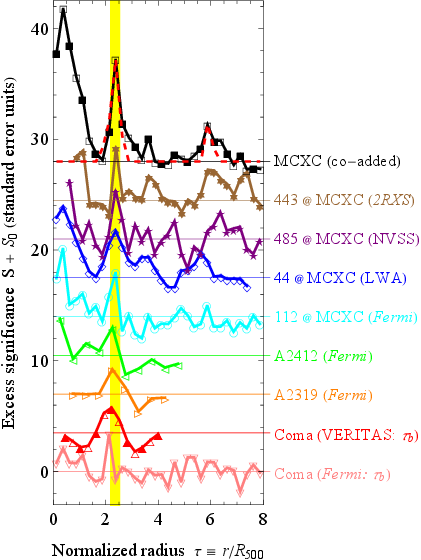}
    \vspace{-0.2cm}}
	\caption{\label{fig:summary}\!\!\!Virial shock signals identified in individual (triangles) or stacked (other symbols) clusters, in diffuse emission (empty symbols), discrete sources (filled), or without separating the two (intermittent empty and filled symbols).
    The significance $S$ (symbols connected by lines to guide the eye, in standard-error units) of the excess above the background $S_0$ (labelled horizontal lines, shifted vertically for visibility) is plotted as a function of the normalized radius $\tau$, for (bottom to top) clusters Coma, in \emph{Fermi}-LAT (down triangles \citep{keshet2018evidence}) and VERITAS (coincident with synchrotron emission and an SZ drop; up triangles \citep{KeshetEtAl17}) data as a function of $\tau_b$, A2319 (right triangles) and A2142 (left triangles) in \emph{Fermi}-LAT data (coincident with SZ drops \citep{keshet20coincident}), and stacked MCXC clusters (labels specify their number) in \emph{Fermi}-LAT (circles \citep{reiss2018detection}) and LWA (diamonds \citep{HouEtAl23}) data and in NVSS (five-pointed stars) and 2XRS (six-stars) source catalogs \citep{IlaniEtAl24a}. Also shown are the $2.2<\tau<2.5$ radial range \citep{reiss2018detection} of MCXC virial-shock signals (vertical yellow band), the co-added MCXC excess (squares), and a corresponding simple cylindrical shock model (dashed red curve; see text). See Table \ref{tab:fig} for more info.
	}
\vspace{-1.1cm}
\end{figure}

Such stacked detections, based on the Meta-Catalog of X-ray detected Clusters of galaxies \citep[MCXC;][]{PiffarettiEtAl11} and at least two additional cluster catalogs \citep{IlaniEtAl24,Nadler24InPrep}, show significant signals narrowly confined near the same normalized $\tau\equiv r/R_{500}\simeq 2.4$ cluster radius in very different channels and data sets, as demonstrated in Fig.~\ref{fig:summary} and detailed in the corresponding Table \ref{tab:fig}.
Similar signals were found in select nearby clusters \citep{KeshetEtAl17, keshet2018evidence, HurierEtAl19, keshet20coincident}, as also illustrated in the figure (triangles).
In addition to the anticipated, smooth synchrotron, inverse-Compton, and Sunyaev-Zel'dovich (SZ) signatures, the figure also shows the surprising coincident excess of discrete radio and X-ray sources, which appear to be galactic halos or galactic outflows energized by the virial shock \citep{IlaniEtAl24a, IlaniEtAl24}.
Note that the detected virial-shock signals roughly coincide with the splashback feature inferred from a localized drop in the logarithmic slopes of galaxy density profiles \citep{MoreEtAl16, ShinEtAl19}.

The aforementioned signals were identified as arising from the virial shock based on their common peripheral location and the agreement of the diffuse-signal properties with predictions, thus constraining the electron acceleration $\xi_e\simeq 1\%$ and magnetization $\xi_B\simeq \mbox{few \%}$ efficiencies, the Mach $\gtrsim 10$ numbers of the shocks, and the $\dot{M}/(MH)\simeq 1$ mass accretion rate onto low-redshift clusters, where $M$ is the cluster mass and $H$ is the Hubble parameter \citep{ReissEtAl17, reiss2018detection, HouEtAl23, keshet2018evidence, HurierEtAl19, keshet20coincident}.
These estimates are based on the diffuse emission remaining after a careful removal of discrete sources, now known \citep{IlaniEtAl24a, IlaniEtAl24} to peak locally at the shock; the residual diffuse radio emission strengthens, as expected \citep{WaxmanLoeb00, KeshetEtAl03}, with $M$ \citep{HouEtAl23}.

Virial shocks are not expected to be spherical.
Hence, the narrow $\tau\simeq 2.4$ excess likely corresponds to a regulated smallest distance from the center of the cluster (the semi-minor axis in an elliptic approximation), whereas a broad, tentative excess around $\tau\simeq 6$ may correspond to the farthest (semi-major) distance \citep{HouEtAl23, IlaniEtAl24a}. Such an interpretation is illustrated in the figure by a simple cylindrical shock model of base radius $\tau=2.4$ and half-height $\tau\simeq6$ (projected, averaged over all orientations, and binned, on top of a fixed background; see \S\ref{sec:Projection}).

\begin{center}
\begin{table}[h!]
\begin{center}
    \vspace{0.3cm}
    \centering
    \begin{center}
    \centering %c@{\hskip 0.4cm}
    {\footnotesize
    \begin{tabularx}{1\linewidth}{l|l|l|l|l}
        Data source & Photon range & Cluster/s & Interpretation & Ref. \\
        \hline
        2RXS & $0.1\mbox{--}2.4\keV$ & MCXC & Galactic halos/outflows & \citenum{IlaniEtAl24a} \\
        NVSS & $1.4\GHz$ & MCXC & Galactic halos/outflows & \citenum{IlaniEtAl24a} \\
        LWA & $73\MHz$ & MCXC & Synchrotron emission & \citenum{HouEtAl23} \\
        \emph{Fermi}-LAT & $1\mbox{--}100\GeV$ & MCXC & Compton emission & \citenum{reiss2018detection} \\
        \emph{Fermi}-LAT & $1\mbox{--}100\GeV$ & A2412 & Compton (+SZ) & \citenum{keshet20coincident} \\
        \emph{Fermi}-LAT & $1\mbox{--}100\GeV$ & A2319 & Compton (+SZ) & \citenum{keshet20coincident} \\
        VERITAS & $\sim220\GeV$ & Coma & Compton (+syn+SZ) & \citenum{KeshetEtAl17} \\
        \emph{Fermi}-LAT & $1\mbox{--}100\GeV$ & Coma & Compton & \citenum{keshet2018evidence} \\
    \end{tabularx}
    }
    \caption{\label{tab:fig}
         \!\!Summary of virial-shock studies demonstrated in Fig.~\ref{fig:summary}, in the same (top to bottom) order. Columns identify the source of the stacked data, the photon energy or frequency range, the cluster analyzed (MCXC when stacked), the suggested interpretation of the signal, and a reference to the analysis. Abbreviations: \emph{Fermi}-LAT --- \emph{Fermi} Large Area Telescope; 2RXS --- Second \emph{ROSAT} All-Sky Survey Source Catalog; NVSS --- National Radio Astronomy Observatory Very Large Array Sky Survey; LWA --- Owens Valley Radio Observatory Long Wavelength Array.
    }
    \end{center}
\end{center}
\end{table}
\end{center}

In addition to the faintness of these virial shock signals, their analysis is complicated by contamination from the ICM.
%intracluster medium (ICM).
Although such peripheral signals can be separated from the central ICM, as demonstrated by the figure, their analysis is thus limited to extended, sufficiently resolved clusters.
Another difficulty is that without individual-cluster imaging, associating a stacked signal with the virial shock relies on projected circumstantial evidence --- radius, flux, and spectrum --- without local support for the directionality of the phenomenon or a measure of the local orientations of the shock and its associated magnetic field.

In order to address these challenges, further establish and map virial shocks, and constrain the shock and magnetic field orientations, we study the polarization of radio emission from stacked clusters, separating its radial and tangential components as defined with respect to the center of the cluster.

Collisionless shocks are thought to generate or amplify magnetic fields parallel to the shock front, so synchrotron emission from shock-accelerated electrons (predicted \citep{WaxmanLoeb00}, simulated \citep{KeshetEtAl03, KeshetEtAl04}, and later detected \citep{KeshetEtAl17,HouEtAl23}) should be polarized perpendicular to the front.
In the idealized case of an approximately spherical virial shock, the edge-on emission should be radially polarized, with a high, $\sim75\%$ polarization fraction in the non-magnetized upstream, strong-shock limit.
Radial polarization is expected from stacked clusters even if the underlying shocks are non-spherical, provided that the signal is well-localized radially, as is the case near $\tau=2.4$.
In particular, this applies to emission from a common minimal or maximal normalized shock radius, where the shock normal is locally radial.
Interestingly, preliminary evidence suggests that the excess $\tau\simeq 2.4$ discrete radio sources indeed tend to be radially polarized \citep{IlaniEtAl24a}.

\section{Data and stacking method}
\label{sec:Methods}

The polarization axis of the emitted radiation is inferred more reliably at high microwave frequencies, where Faraday rotation by the Galactic magnetic field becomes negligible.
However, the stacked synchrotron virial-shock signal is weak \citep{HouEtAl23}, and is thought to diminish at such high frequencies with respect to the Galactic foreground and extragalactic background \citep{KeshetEtAl04, KeshetEtAl17}.
Indeed, we are unable to pick up an MCXC-stacked virial-shock signal even at the lowest, $30\GHz$ \emph{Planck} frequency.
We thus use the Global Magneto-Ionic Medium Survey (GMIMS) high-band (1280 to 1750 MHz) data release \citep{WollebenEtAl21}, covering the northern sky at declinations $-30\dgr<\delta<87\dgr$ observed using the John A. Galt Telescope at the Dominion Radio Astrophysical Observatory (DRAO).
Although Faraday rotation in GMIMS is prohibitively strong near the Galactic plane, there is a high-latitude window where it becomes negligible.

In particular, at high, $|b|>70\dgr$ latitudes, the mean Galactic Faraday rotation measure drops to $\mbox{RM}\lesssim (10\pm20)\mbox{ rad m}^{-2}$ \citep{TaylorEtAl09}.
Sky coverage varies among GMIMS channels, but is uniformly maximal in the high, $[\nu_l,\nu_h]\equiv[1630.74,1734.94]\MHz$ frequency band, which we utilize in the present analysis.
At these frequencies, the $\mbox{RM}(|b|>70\dgr)$ values above correspond to small, $\chi(\nu_h)\lesssim17\dgr\pm34\dgr$ rotation angles, so the radial and tangential polarizations can be distinguished even without correcting for Faraday rotation.
Moreover, the high-latitude RM is $\sim$twice smaller in the northern hemisphere, and is vanishingly small near the north Galactic pole \citep{TaylorEtAl09}, where $\mbox{RM}(b>75\dgr)\simeq 3.0\pm0.5\mbox{ rad m}^{-2}$ translates to only $\chi(\nu_h)\simeq 5\dgr\pm1\dgr$.

GMIMS Stokes $Q$ and $U$ HEALPix (Hierarchical Equal Area isoLatitude Pixelization) maps \footnote{With polarization angle in IAU conventions (T. Landecker, private communications): counter-clockwise on the plane of the sky, north due east.} at resolution index 8 are used, with a number $N_{\mbox{\tiny{side}}}=256$ of pixels per side implying 786,432 all-sky pixels with a $\sim0\dgrdot23$ angular separation and a fixed solid angle $\delta\Omega\simeq 1.6\times 10^{-5}\sr$.
These maps are stacked around MCXC clusters, scaled to their $\theta_{500}\equiv R_{500}/d_A(z)$ characteristic angles,
after separating the total, $P=(Q^2+U^2)^{1/2}$ polarization into radial and tangential polarized components, $P_r$ and $P_t$, as a function of the angle \footnote{The angle $\phi$ defined with polarization-angle conventions: north due east.} $\phi(\Myc,\MyH)$ of a given HEALPix pixel $\MyH$ with respect to the center of nearby cluster $\Myc$.
Here, $d_A(z)$ is the angular diameter distance at the cluster redshift $z$.

We adopt the beam co-addition stacking method, already established for radio virial-shock signals \citep{HouEtAl23}, as follows.
For each channel $\Myfr$, cluster $\Myc$, and pixel $\MyH$, denote the measured
brightness temperature as $T(\Myfr,\Myc,\MyH)$, and the corresponding sum of background and foreground temperatures (henceforth background, for brevity) as $\MyT(\Myfr,\Myc,\MyH)$, so the local excess becomes $\Delta T(\Myfr,\Myc,\MyH) \equiv T-\MyT$.
This procedure is repeated for each polarization component.
Namely, for total, radial, and tangential polarizations, $\Delta T$ becomes $\Delta P\equiv P-P_b$, $\Delta P_r\equiv \Delta P |\cos\delta\phi|$, and $\Delta P_t\equiv\Delta P |\sin\delta\phi|$, respectively, where the angle difference
\begin{equation}\label{eq:dphi}
\delta \phi(\Myfr,\Myc,\MyH) \equiv \frac{\arg(\Delta Q+i\,\Delta U)}{2}-\phi(\Myc,\MyH)
\end{equation}
is measured between local cluster radius and polarization axis.
The excess brightness profile, averaged over a sample of $N_c$ clusters, now becomes
\begin{equation}
\label{eq:ExcessInu}
\Delta I_{\Myfr}(\Myfr, \tau)
=
\frac{2\nu^2}{c^2}\frac{\sum_{\Myc=1}^{N_c} \sum_{\MyH=1}^{N_p} k_B\Delta T(\Myfr,\Myc,\MyH)}{\sum_{\Myc=1}^{N_c} N_p(\tau,\Myc) } \coma
\end{equation}
where $k_B$ is the Boltzmann constant, $c$ is the speed of light, and
$N_p (\tau,\Myc)$ is the number of HEALPix pixels falling in the radial $\tau$ bin of cluster $\Myc$.

Background fields $T_b$ are nominally defined for each cluster as constants, given by the average in the $5<\tau<15$ radial range.
Different choices of background reference region are examined, as well as replacing the uniform background by linear or quadratic $T_b(\tau_x,\tau_y)$ fits around each cluster on the plane of the sky, with no qualitative change in the results.
The brightness profile \eqref{eq:ExcessInu} is nominally binned at resolution $\Delta\tau=1/4$, but larger and smaller bins are also inspected.
Brightness profiles are examined both in individual frequencies, and co-added over the entire $[\nu_l,\nu_h]$ band.
In the latter case, we average the frequency-weighted $\nu\,\Delta I_\nu$, in anticipation of a spectrally flat virial-shock synchrotron signal \citep{HouEtAl23}, but such weights make very little difference given the present narrow frequency range.

The significance $S$ of the excess brightness, stacked over $N_c$ clusters, may be estimated in standard error units as
\begin{equation}\label{eq:BeamSig}
\MySig(\nu,\tau) = \frac
{\sum_{\Myc=1}^{N_c} \sum_{\MyH=1}^{N_p} \Delta T(\Myfr,\Myc,\MyH)}
{\eta T_0 \sqrt{\pxN\sum_{\Myc=1}^{N_c} N_p(\tau,\Myc)}} \coma
\end{equation}
where
\begin{equation} \label{eq:GMIMS_Beam}
\pxN
\simeq \frac{\pi\left( \mbox{FWHM}/2\right)^2}{\delta\Omega}
\simeq 3.7
\end{equation}
is the number of (correlated, HEALPix) pixels in the beam, and we adopted \citep{WollebenEtAl21} the full-width half maximum $\mbox{FWHM}=30'$.
Here, we assumed that the noise is predominantly thermal, at the $T_0\simeq 45\mK$ noise temperature (for $Q$ and $U$; see \citep{WollebenEtAl21}), and given by normal statistics among the
$\sim N_p/\pxN$
independent beams mapped onto
the $\tau$ bin.
The order-unity correction factor $\eta$ is introduced because the product $T_0\cdot\mbox{FWHM}$ is not precisely known, and to mitigate putative biases and correlations on the sky.
When co-adding $N_\nu$ frequency channels, $S(\tau)=N_\nu^{-(1+\psi)/2}\sum_\nu  S(\nu,\tau)$, we introduce an additional $0<\psi<1$ parameter to take into account the anticipated strong \citep{HouEtAl23,WollebenEtAl21} inter-channel correlations.

In order to examine the statistical properties of the background, calibrate $\eta$ and $\psi$, test the normality approximation underlying Eq.~\eqref{eq:BeamSig}, and substantiate this $S$ estimate, we analyze $>4000$ Monte-Carlo control cluster samples for each MCXC sample studied.
Each such control sample is similar to the true cluster sample, using in particular the same number of clusters with the same corresponding MCXC redshift and $R_{500}$ values, but the coordinates of each control cluster are randomized over the relevant sky region.

Analyzing single-channel control samples (see Table \ref{tab:eta} and \S\ref{sec:Results}) indicates that (i) the background statistics are approximately normal on the relevant, $1<\tau\lesssim 15$ scales; (ii) for our parameters, $\eta\simeq 0.3$ near the north Galactic pole, with no strong variations with sky position, resolution, or frequency; and (iii) the estimate \eqref{eq:BeamSig} is reliable within $\sim10\%$ accuracy near the virial shock.
Frequency co-added control samples indicate that $\psi\simeq 0.8$, consistent with the LWA stacking analysis (after restoring point sources with a Gaussian beam \citep{HouEtAl23}), confirming strong correlations among channels.

\begin{center}
\begin{table}[h!]
\begin{center}
    \vspace{0.3cm}
    \centering
    \begin{center}
    \centering %c@{\hskip 0.4cm}
    \begin{tabularx}{1\linewidth}{X|XX|XX|XX|XX}
        $\Delta\tau$ & \multicolumn{2}{c|}{$\eta_P$} & \multicolumn{2}{c|}{$\eta_r$} & \multicolumn{2}{c|}{$\eta_t$} & \multicolumn{2}{c}{$\psi$} \\
        & VS & ROI & VS & ROI & VS & ROI & VS & ROI \\
        \hline
        $1/4$ & 0.32 & 0.30 & 0.30 & 0.28 & 0.30 & 0.29 & 0.80 & 0.79 \\
        $1/8$ & 0.27 & 0.26 & 0.26 & 0.25 & 0.26 & 0.25 & 0.77 & 0.77
    \end{tabularx}
    \caption{\label{tab:eta}
         \!\!Stacking parameter calibration using $b>70\dgr$ control samples in the virial-shock vicinity (VS; $2<\tau<3$) and in the cluster region of interest (ROI; $0<\tau<5$).
        Beam correction factors $\eta$ of Eq.~\eqref{eq:BeamSig} are provided for total ($\eta_P$), radial ($\eta_r$), and tangential ($\eta_t$) polarized components, in nominal and high ($\Delta\tau=1/8$) resolutions. Channel-correlation factors $\psi$ are consistent across these polarizations, within $1\%$.
    }
    \end{center}
\end{center}
\end{table}
\end{center}

Although source catalogs (like NVSS \citep{NVSS_paper} and Rapid ASKAP Continuum Survey \citep[RACS;][]{DuchesneEtAl24}) are available at relevant frequencies, they contain more sources than GMIMS HEALPix pixels, so individual sources cannot be removed or masked.
However, as the control samples quantify the properties of the GMIMS sky, the $\eta$-corrected $S$ estimate already accounts for the statistical properties of embedded point sources, as well as any other sky components and correlations.
Any virial-shock signal should however be considered a combination of diffuse and discrete-source emission, given previous detections of a strong source excess at the shock.
The anticipated $P_r$ signal is expected to be somewhat weakened by noise bias, due to $Q$ and $U$ noise cross terms, and (in the absence of a non-polarized excess; see \S\ref{sec:Discussion}) by polarization leakage.
Another nonlinear bias arises from the removal of background fields in both $P$ and $\delta\phi$ (which depends on $\Delta Q$ and $\Delta U$); to test this bias, we also consider an alternative
\begin{equation}\label{eq:PTilde}
\Delta P = \sqrt{\Delta Q^2 + \Delta U^2}\coma
\end{equation}
found to produce consistent but more noisy results.

\section{Radially polarized excess}
\label{sec:Results}

\begin{figure}[h!]
    \centerline{
        \includegraphics[width=1\linewidth,trim={0cm 1.6cm 0cm 0cm},clip]{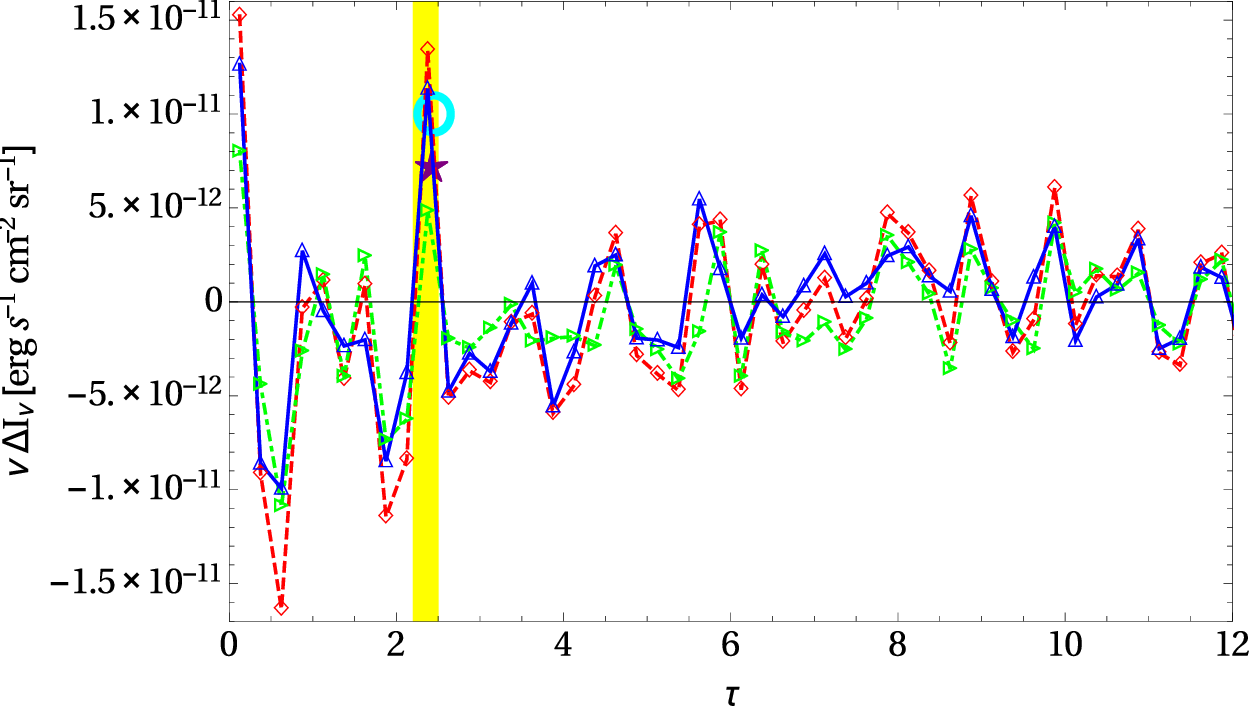}
    }
    \centerline{
        \hspace{0.95cm}\includegraphics[width=0.892\linewidth]{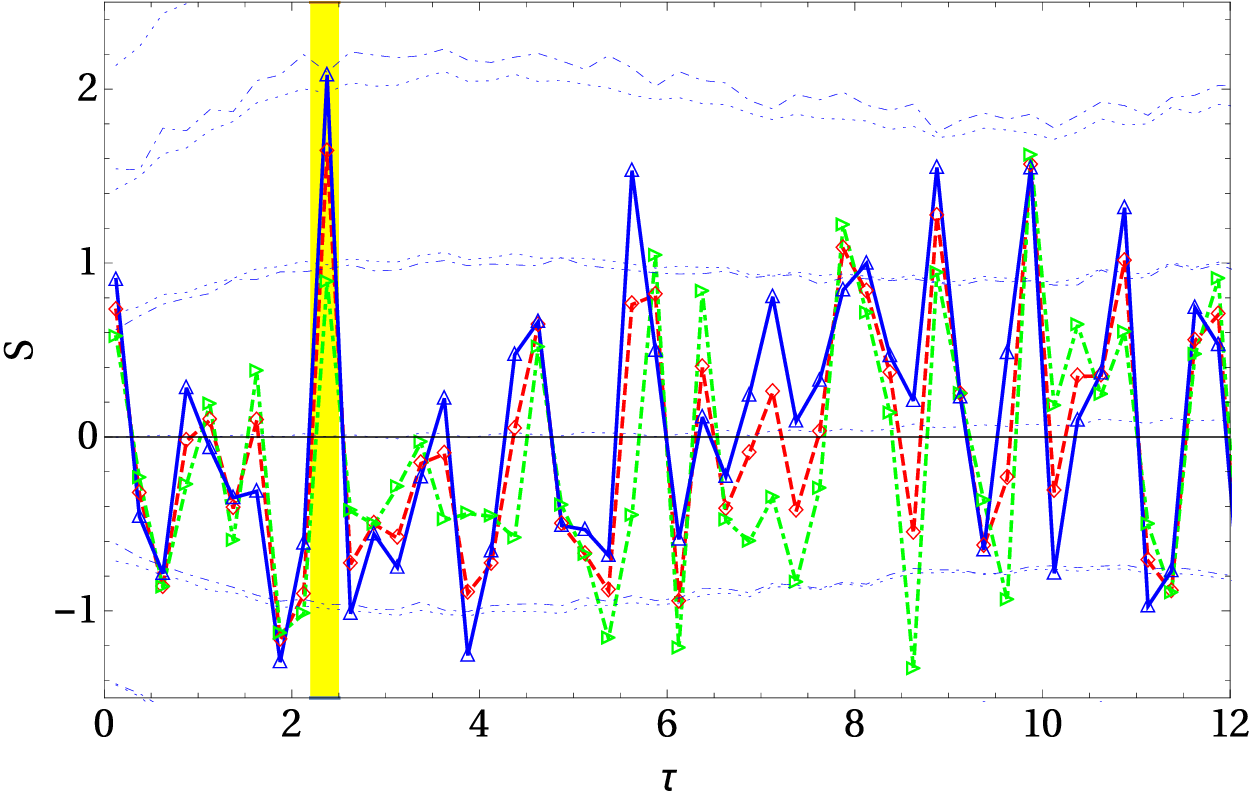}
    }
	\caption{\label{fig:nuIh}
    Total (red diamonds with dashed lines to guide the eye), radial (blue up triangles, solid guides), and tangential (green right triangles, dot-dashed guides) polarized GMIMS excess above the respective background, stacked over the nominal ($M_{500}>10^{14}M_\odot$, $b>70\dgr$) 85-cluster sample, averaged within concentric $\tau$ bins in the high, $\nu_h$ frequency channel. The virial-shock radial range of previous MCXC signals is highlighted (vertical yellow band).\\
    %Three highly extended, $\theta_{500}>0\dgrdot5$ clusters were excluded. \\
    Top panel: Mean cluster excess-brightness according to Eq.~\eqref{eq:ExcessInu}. Shown for reference are non-polarized radio virial-shock excess signals, in diffuse emission extrapolated from LWA frequencies (\citep{HouEtAl23}, assuming a flat spectrum; cyan circle) and in NVSS sources (\citep{IlaniEtAl24a}; purple star); see \S\ref{sec:Discussion}. \\
    Bottom panel: Significance of the local excess in standard error units, according to Eq.~\eqref{eq:BeamSig}. Also shown are control-sample $n=0,\pm1,\pm2,\ldots$ confidence levels, as containment brackets (for small $|n|$; dot-dashed blue curves) and in the Gaussian approximation ($\mu+n\sigma$; dotted blue).
	}
\end{figure}

The preceding discussion indicates that radially polarized synchrotron emission from virial shocks would be most noticeable around high-latitude, massive clusters in the northern hemisphere, at high frequencies where Faraday rotation weakens.
Our nominal cluster selection thus consists of the 85 high, $b>70\dgr$ latitude, massive, $M_{500}>10^{14}M_\odot$ MCXC clusters; median parameters are $b\simeq 77\dgr$, $M_{500}\simeq 2.1\times 10^{14}M_\odot$, and $z\simeq 0.21$.
Figure \ref{fig:nuIh} shows GMIMS northern sky data in the high, $\nu_h$ frequency channel, without any RM correction, stacked around these clusters after scaling to their respective $\theta_{500}$ radii.

Earlier MCXC stacking analyses, outlined in \S\ref{sec:Intro}, had to exclude compact clusters of small $\theta_{500}$ in order to avoid contaminating the virial radius with strong central-signal residuals, but there is no need to do so in the present case, as the ICM is not particularly bright in radially polarized emission, and the very central cluster regions should be highly depolarized.
We do however exclude three highly extended, $\theta_{500}>0\dgrdot5$ clusters (Coma being one of them), in order to avoid their many pixels from dominating the stacked signal; the outcome is not sensitive to the precise upper limit on $\theta_{500}$.

The top panel of Fig.~\ref{fig:nuIh} shows the radial $\nu \,\Delta I_\nu(\tau)$ profile of the excess in stacked brightness, in total, radial, and tangential polarizations, above their respective nominal (based on $5<\tau<15$) backgrounds, according to Eq.~\eqref{fig:nuIh}.
The bottom panel shows the corresponding local significance $S(\tau)$ profile of each of these excess signals, as estimated from the beam co-addition (ordinate) of Eq.~\eqref{eq:BeamSig}, and in comparison to the Monte-Carlo control samples (dotted and dot-dashed blue curves).
A nominal $\sim2\sigma$ (with respect to the $5<\tau<15$ background) but local $\sim3\sigma$ (with respect to the local background in nearby bins) excess in radially polarized emission is seen around $\tau\simeq 2.4$, in the exact same radial range (vertical yellow band) of other virial-shock signals found in previous MCXC stacking analyses.
Note that those studies have not focused on such high and northern latitudes, so did not rely on the same specific clusters used here.
As expected, the excess arises mostly from the 20 more extended, $\theta_{500}>0\dgrdot1$ clusters (median parameters $b\simeq 76\dgr$, $M_{500}\simeq 1.9\times 10^{14}M_\odot$, and $z\simeq 0.076$), although the 65 more compact clusters make a noticeable contribution, too.

\begin{figure}[h!]
    \centerline{
        \includegraphics[width=1\linewidth,trim={0cm 0.8cm 0cm 0cm},clip]{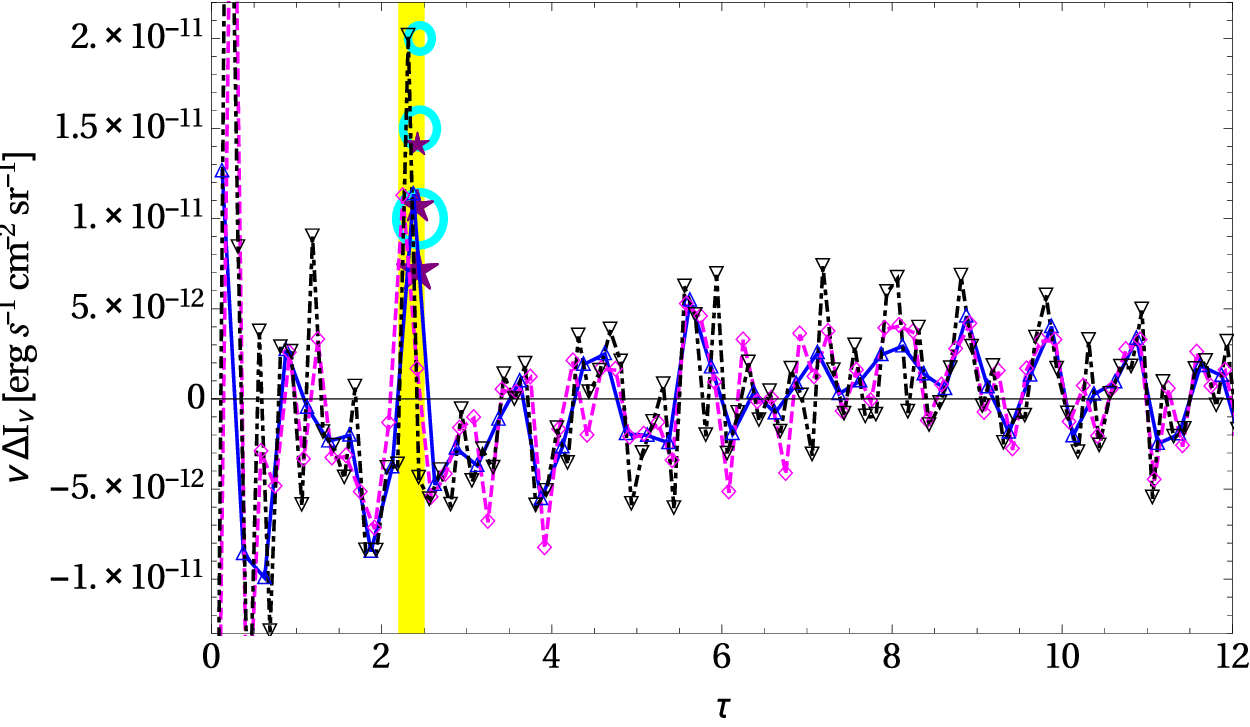}
    }
    \centerline{
        \hspace{0.82cm}\includegraphics[width=0.91\linewidth,trim={0cm 0.8cm 0cm 0cm},clip]{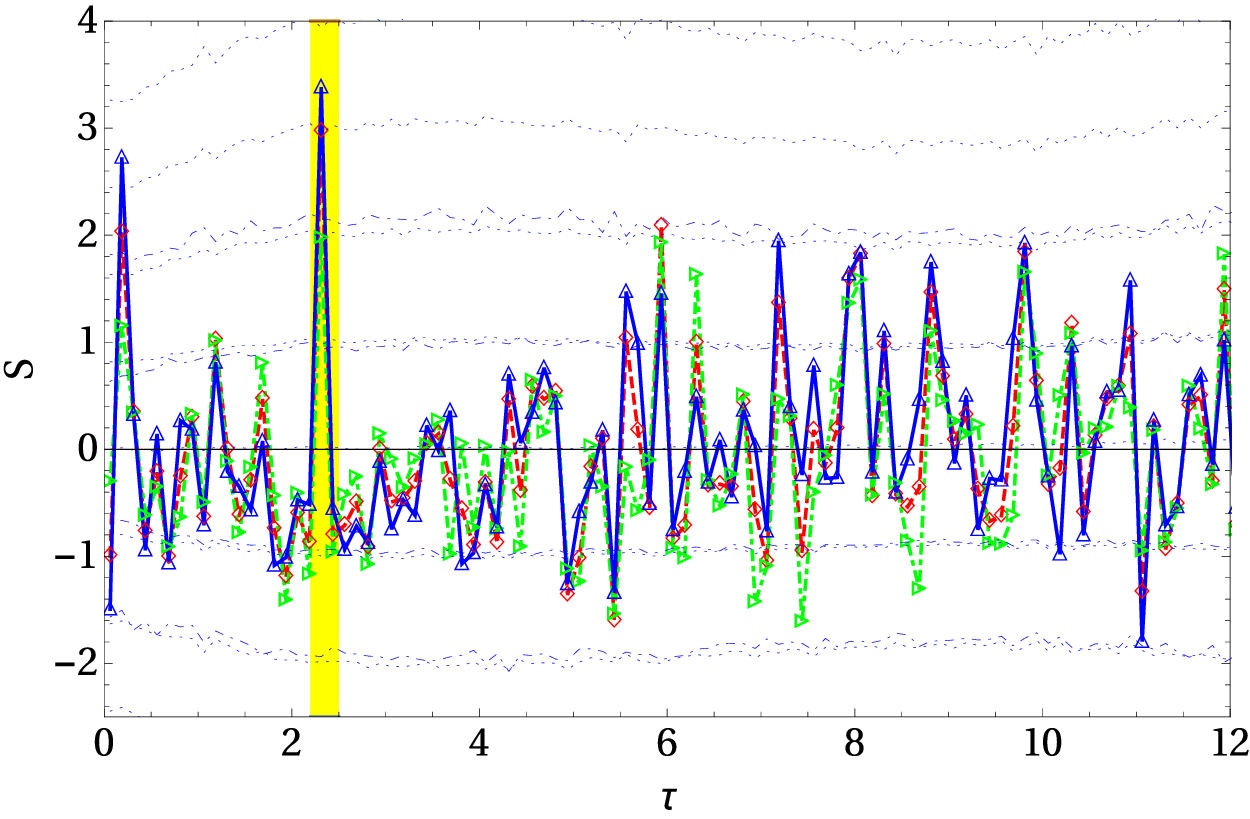}
    }
    \centerline{
        \includegraphics[width=1\linewidth,trim={0cm 0.8cm 0cm 0cm},clip]{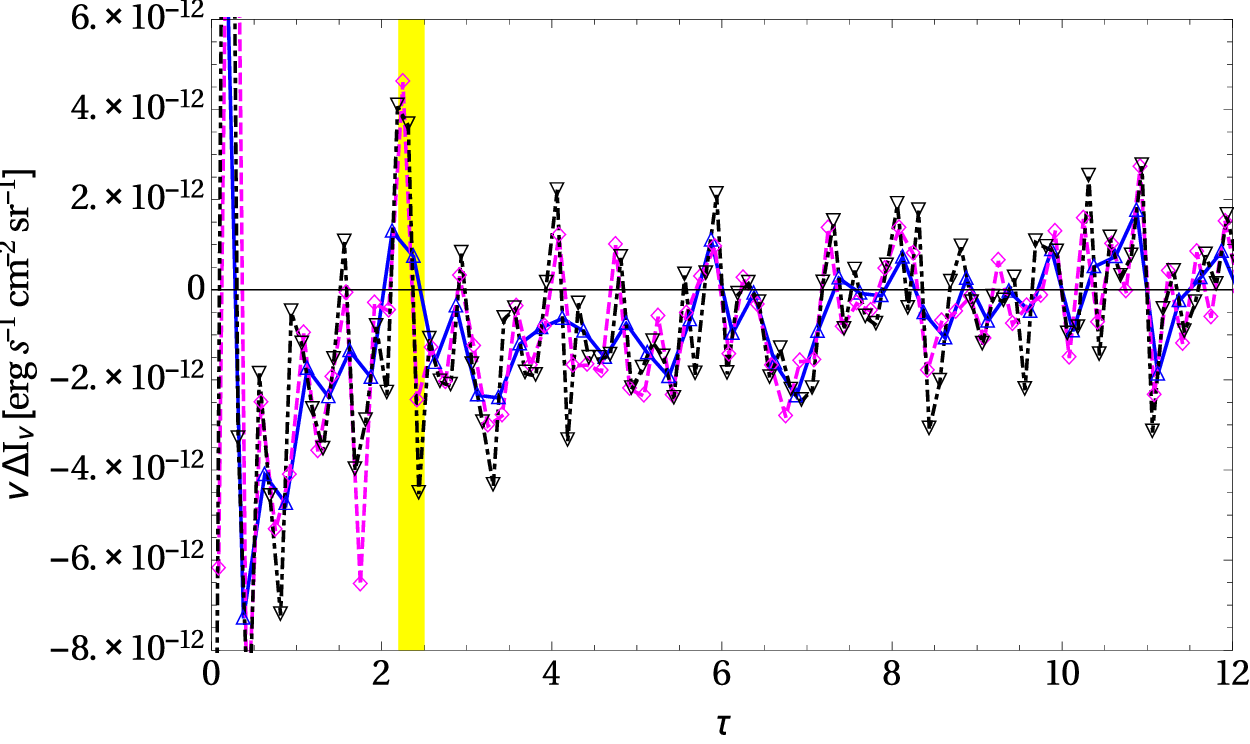}
    }
	\caption{\label{fig:nuIhTauRes}
    Nominal cluster sample stacked as in Fig.~\ref{fig:nuIh}, but with different resolutions and frequencies.\\
    Top panel: Radial polarization in the $\nu_h$ channel for resolutions $\Delta \tau=1/4$ (nominal; blue up triangles with solid guide), $\Delta\tau=1/6$ (magenta diamonds; dashed guide), and $\Delta\tau=1/8$ (black down triangles; dash-dotted guide), each with its corresponding reference (large to small symbols) non-polarized LWA-extrapolated diffuse (cyan circle) and NVSS source (purple star) emission.\\
    Middle panel: Same as the bottom panel of Fig.~\ref{fig:nuIh}, but for $\Delta\tau=1/8$ resolution.\\
    Bottom panel: Same as the top panel, but averaged over $[\nu_l,\nu_h]$ frequencies.\\
	}
\end{figure}

\begin{figure}[h]
    \centerline{
        \includegraphics[width=1\linewidth]{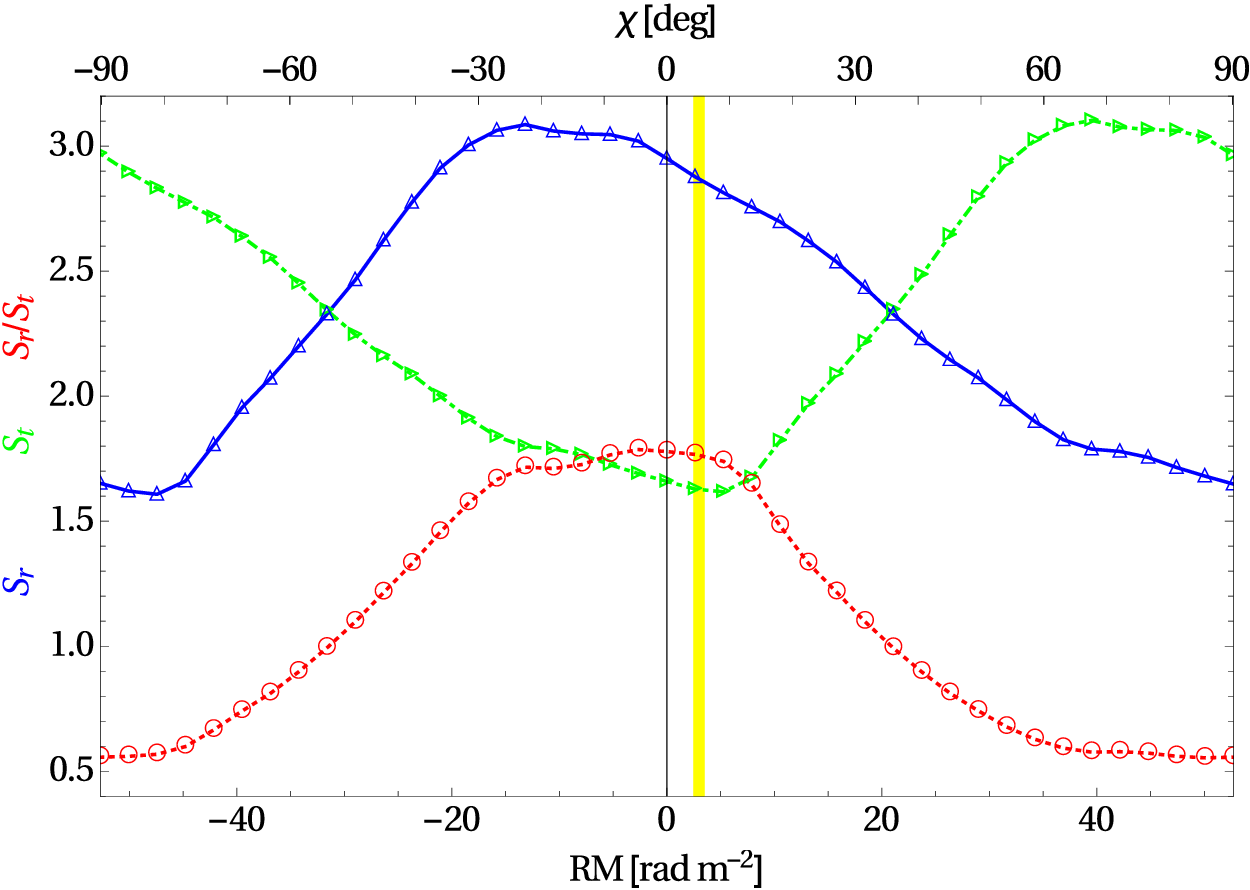}
    }
	\caption{\label{fig:RM}
        Significance of the nominal local excess of radial ($S_r$; blue up triangles) and tangential ($S_t$; green right triangles) polarizations in the virial-shock bin, and their ratio ($S_r/S_t$; red circles), after correcting for different putative Galactic Faraday RM values (abscissa; top axis shows the corresponding rotation angle) in the $\nu_h$ channel.
        Also shown is the measured \citep{TaylorEtAl09} $\mbox{RM}=(3.0\pm0.5)\mbox{ rad m}^{-2}$ range (for $b>75\dgr$; vertical yellow band).
        Sign conventions are standard: $\mbox{RM}>0$ implies that magnetic fields, preferentially oriented toward the observer, have rotated the polarization vector counter-clockwise on the sky.
	}
\end{figure}

We find a similar excess of radially polarized emission in other GMIMS channels and when varying the parameters of the analysis.
Figure \ref{fig:nuIhTauRes} demonstrates that the signal strengthens at higher resolutions, and remains present even when co-adding all 89 channels in the $[\nu_l,\nu_h]$ range.
The persistence of the signal upon channel co-addition confirms that Faraday rotation is weak near the north Galactic pole.
Indeed, Fig.~\ref{fig:RM} shows that the radial-to-tangential polarization ratio is maximal if one assumes $\mbox{RM}\simeq 0$, consistent with preferentially radial emission undergoing negligible Faraday rotation.
Co-adding channels does not in general raise the signal-to-noise ratio, consistent with channel-correlated sky noise, as indicated by the large $\psi$.

Strong confinement of the radially binned signal near $\tau\simeq 2.4$ was previously identified in some other virial-shock tracers \citep{reiss2018detection, keshet20coincident, IlaniEtAl24a, IlaniEtAl24}, despite stacking, projection effects, and limited resolution, localizing the stacked signal within a radial range narrower than the two-dimensional FWHM, supporting a universal extremal $\tau$.
Polarized synchrotron emission is anticipated mainly where the virial shock is seen edge-on, so the present projected signal is expected to be even narrower than found in its non-polarized counterparts.
Indeed, the excess in Fig.~\ref{fig:nuIhTauRes} is confined within a very narrow radial range; the local significance of the signal exceeds $4\sigma$ when the nominal resolution is doubled to $\Delta\tau=1/8$.
The nominal FWHM of GMIMS, smaller than $\theta_{500}$ in our sample, may have been overestimated, as the underlying scans have a higher resolution \citep{WollebenEtAl21}; indeed, the $\eta$ values obtained here are small.

Modeling the signal as a thin ring on the sky (the planar model of Refs.~\citep{keshet20coincident, HouEtAl23, IlaniEtAl24a}) yields TS-test values of $5.8$ for $\Delta\tau=1/4$ resolution, and $13.4$ for $\Delta\tau=1/8$.
Here, $\mbox{TS} = \chi^2_- - \chi^2_+$ compares $\chi^2$ values obtained by fitting models of $\mathsf{n}$ parameters, obtained before ($-$ subscripts) and after ($+$ subscript) supplementing the background with a virial-shock component; TS then approximately follows a chi-squared distribution $\chi_\mathsf{n}^2$ of order $\mathsf{n}\equiv \mathsf{n}_+-\mathsf{n}_-$ \citep{Wilks1938}.
As the shock radius was already pre-determined, these one-added-parameter, $\mathsf{n}=1$ models correspond to virial-shock detections at the $2.4\sigma$ and $3.7\sigma$ confidence levels, respectively.

\begin{figure}[h]
    \centerline{
        \includegraphics[width=1\linewidth]{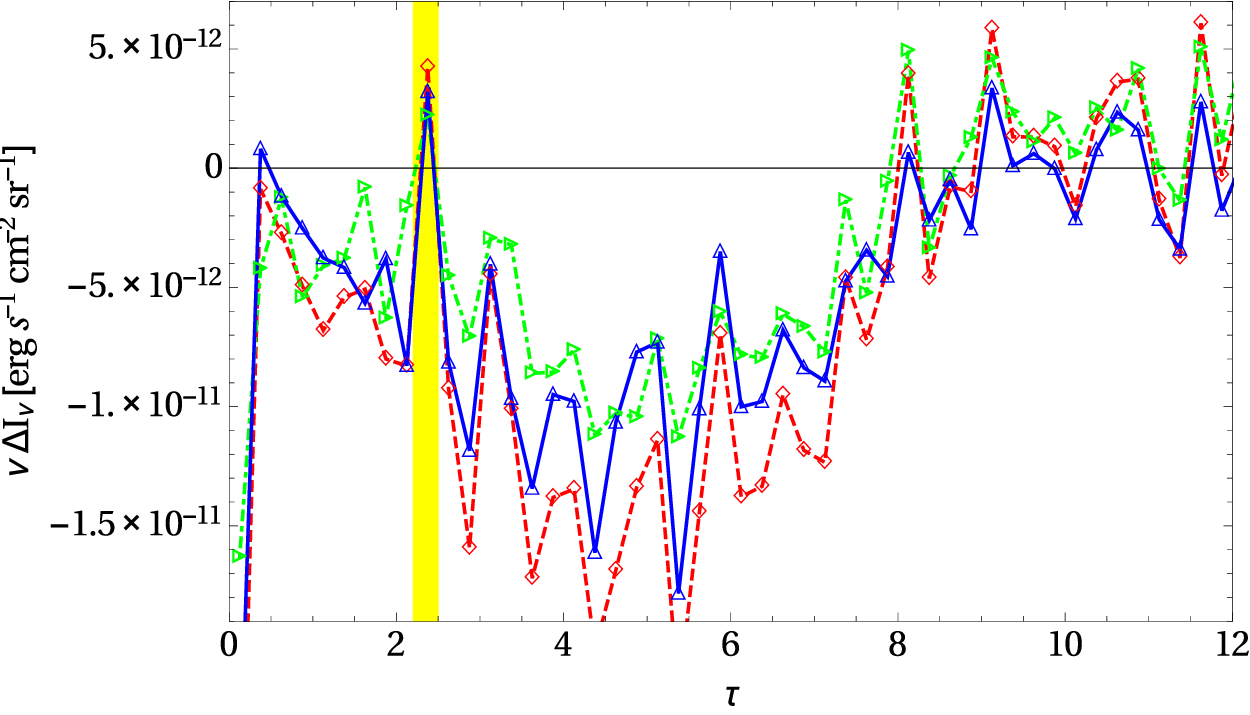}
    }
	\caption{\label{fig:LowM}
        Same as Fig.~\ref{fig:nuIh} (top panel), but for a sample of lower, $10^{13}<M_{500}/M_\odot<10^{14}$ mass clusters and groups.
	}
\end{figure}

Synchrotron emission from virial shocks should be a strong, $I_\nu\propto \dot{M}TB^2/r_s^2\propto M^{5/3}$ function of cluster mass, and the LWA signal is indeed noticeably weaker for $M_{500}<10^{14}M_\odot$ clusters \citep{HouEtAl23}.
Here, we used the scaling arguments of Refs.~\citep{KeshetEtAl04, HouEtAl23}, where $\dot{M}$, $T$, $B$, and $r_s$ are respectively the cluster mass accretion rate, temperature, post-shock magnetic field, and shock radius.
Figure \ref{fig:LowM} shows that, as expected, the radially polarized signal weakens for low, $10^{13}<M_{500}/M_\odot<10^{14}$ cluster masses, although it is still noticeable.
However, as there are only 42 clusters in this mass range (for the same nominal $b>70\dgr$ and $\theta_{500}<0\dgrdot5$ cuts, giving a sample of median parameters $b\simeq 76\dgr$, $M_{500}\simeq 5.7\times 10^{13}M_\odot$, and $z\simeq 0.13$), we cannot quantify this effect and test the $5/3$ power law index.

Stacked GMIMS signals in the southern hemisphere show prohibitively strong fluctuations, preventing us from reaching the sensitivity needed to detect the signal near the south Galactic pole.
These stronger southern fluctuations appear to arise mainly from the poor statistics of southern MCXC clusters in the northern GMIMS survey field of view.
For instance, there are only 17 (only 28) massive, $M_{500}>10^{14}M_\odot$ clusters meeting our criteria in the south, $b<-70\dgr$ region, if we avoid low, $\delta<-20\dgr$ ($\delta<-25\dgr$) declinations approaching the edge of the survey.
An additional difficulty is that the high $|b|$ RM is more than twice larger in the south \citep{TaylorEtAl09}.

\section{Discussion}
\label{sec:Discussion}

Figures \ref{fig:nuIh} and \ref{fig:nuIhTauRes}, along with the corresponding control samples and TS-tests, show a significant, $3\sigma$--$4\sigma$ local excess of radially polarized radio emission around massive high-latitude clusters, originating from the very same $\tau\simeq 2.4$ scaled cluster-radii of previously identified virial-shock signals (\emph{cf.} Fig.~\ref{fig:summary}).
The results are consistent with the high degree of radial localization (Fig.~\ref{fig:nuIhTauRes}), negligible Faraday rotation (Figs.~\ref{fig:nuIhTauRes} and \ref{fig:RM}), and strong mass dependence (Fig.~\ref{fig:LowM}), anticipated in this part of the sky for synchrotron emission from virial shock-accelerated electrons, as they gyrate in the magnetic fields induced mainly parallel to the shock front.
A coincident excess can also be seen in total polarization, but is less significant than in the radial component alone, while the tangentially polarized excess is noticeably weaker and can be attributed to the dispersion in magnetic field directions and the non-linear bias (see \S\ref{sec:Methods}).

Determining the polarization fraction is difficult, as strong background fluctuations preclude a measurement of the coincident non-polarized synchrotron signal.
However, the LWA non-polarized $\nu I_\nu(\nu\simeq73\MHz) \simeq 10^{-11}\erg\se^{-1}\cm^{-2}\sr^{-1}$ virial-shock brightness (\citep{HouEtAl23}; for $\Delta\tau=0.25$) can be extrapolated to GMIMS frequencies by assuming a flat (constant $\nu I_\nu$) spectrum, as expected from such strong shocks  and consistent with inverse-Compton \citep{keshet2018evidence, reiss2018detection}, synchrotron \citep{HouEtAl23}, and SZ \citep{keshet20coincident} signals.
This extrapolated brightness (cyan circles in Figs.~\ref{fig:nuIh} and \ref{fig:nuIhTauRes}) is comparable to the polarized signal, implying both a high polarization fraction and a flat spectrum, assuming that shock acceleration cannot produce any harder spectra.
The spectrum is thought to be slightly softer than flat, thus raising the inferred polarization fraction.
However, this fraction cannot be determined with an accuracy better than a factor of a few, as the spectrum is not well-determined, the frequency extrapolation is substantial, and the clusters stacked in LWA and GMIMS analyses differ.

Overall, the results support previous claims that a well localized signal emerges around $2.4R_{500}$ in different channels and data sets, in particular in synchrotron emission \citep{KeshetEtAl17, HouEtAl23},
provide the first directional evidence for this signal, as the radial polarization implies transverse magnetization, and thus also verifies that the coincident magnetization is driven by a shock.
When combined with the localization of other signals, the radial polarization also indicates a locally transverse shock geometry, which agrees with $2.4 R_{500}$ being a minimal shock radius, consistent with a simple cylindrical geometry evaluated in \ref{sec:Projection} and demonstrated in Fig.~\ref{fig:summary}.
The properties of the signal are consistent with the expected spectrum and mass dependence of virial-shock emission.
Implications for collisionless shock physics are deferred to Ref.~\citep{KeshetHou24}.

Discrete sources powered by the virial shock \citep{IlaniEtAl24a, IlaniEtAl24} should also be taken into account, as their highly localized emission contributes to the stacked virial excess, and compact sources were not removed from the GMIMS data.
Approximating such sources as having a characteristic $\sim 130\mbox{ mJy}$ NVSS flux each, and an average abundance of one source per $\sim 12$ clusters (\citep{IlaniEtAl24a}; at $z\lesssim0.2$ redshifts), corresponds to a mean non-polarized $\nu F_\nu\simeq 1.6\times 10^{-16} \erg\se^{-1}\cm^{-2}$ flux per cluster.
The implied mean brightness (purple stars in Figs.~\ref{fig:nuIh} and \ref{fig:nuIhTauRes}) carries an uncertainty factor of order a few, due to the flux cut imposed on the NVSS stacking analysis, the highly uncertain flux of individual NVSS sources contributing to the virial excess, and the different clusters stacked when analyzing NVSS vs. GMIMS data.
This non-polarized estimate is again comparable to the GMIMS polarized signal, and the excess NVSS sources do show evidence for a predominantly radial polarization.
However, the polarization fractions of these sources are small, in the $1\%$--$10\%$ range \citep{IlaniEtAl24a}, suggesting that the GMIMS signal is dominated by diffuse emission and not by discrete sources.

Notably, total-polarization signals in GMIMS and \emph{Planck} data, stacked around $\sim6\times 10^5$ galaxy filaments and low-mass, $M<10^{14}M_\odot$ group/cluster candidates, were recently claimed to be associated with accretion shocks \citep{VernstromEtAl23}.
However, these signals arise from small radii, and are unlikely to be directly related to accretion shocks: the GMIMS signal is centrally bright, peaking at $<0.5\Mpc$ radii, and the \emph{Planck} signal is dominated by a relatively small, $r\simeq 1\Mpc$ ring after central depolarization.
Indeed, stacking on physical scales, without normalizing by the characteristic $R_{500}$ radii of each cluster, was generally found to have insufficient sensitivity to the weak virial shock signals, at least for clusters in the relevant low ($z\lesssim0.2$) redshifts (see \S\ref{sec:Intro}).
Rather, these total-polarization ICM signals are more likely to arise from a population of radio relics associated with weaker, merger or similar shocks in the ICM \citep{VernstromEtAl23}.
Such relics are probably related to virial shocks, but only indirectly, as evidence suggests that the radiating electrons are secondaries produced by virial-shock accelerated ions \citep{Keshet10, Keshet24, Keshet25Phoenix, Keshet25PaperIII} (thus challenging the standard lore of primary electron acceleration, \emph{e.g.} \citep{vanWeerenEtAl19} and references therein).

\acknowledgments
I am grateful to K.~C. Hou, G. Nadler, Y. Lyubarsky, and the late G. Ilani, for inspiring discussions.
This research was supported by the Israel Science Foundation (ISF grant No. 2126/22).

\bibliography{Virial}

\begin{thebibliography}{40}%
\makeatletter
\providecommand \@ifxundefined [1]{%
 \@ifx{#1\undefined}
}%
\providecommand \@ifnum [1]{%
 \ifnum #1\expandafter \@firstoftwo
 \else \expandafter \@secondoftwo
 \fi
}%
\providecommand \@ifx [1]{%
 \ifx #1\expandafter \@firstoftwo
 \else \expandafter \@secondoftwo
 \fi
}%
\providecommand \natexlab [1]{#1}%
\providecommand \enquote  [1]{``#1''}%
\providecommand \bibnamefont  [1]{#1}%
\providecommand \bibfnamefont [1]{#1}%
\providecommand \citenamefont [1]{#1}%
\providecommand \href@noop [0]{\@secondoftwo}%
\providecommand \href [0]{\begingroup \@sanitize@url \@href}%
\providecommand \@href[1]{\@@startlink{#1}\@@href}%
\providecommand \@@href[1]{\endgroup#1\@@endlink}%
\providecommand \@sanitize@url [0]{\catcode `\\12\catcode `\$12\catcode
  `\&12\catcode `\#12\catcode `\^12\catcode `\_12\catcode `\%12\relax}%
\providecommand \@@startlink[1]{}%
\providecommand \@@endlink[0]{}%
\providecommand \url  [0]{\begingroup\@sanitize@url \@url }%
\providecommand \@url [1]{\endgroup\@href {#1}{\urlprefix }}%
\providecommand \urlprefix  [0]{URL }%
\providecommand \Eprint [0]{\href }%
\providecommand \doibase [0]{https://doi.org/}%
\providecommand \selectlanguage [0]{\@gobble}%
\providecommand \bibinfo  [0]{\@secondoftwo}%
\providecommand \bibfield  [0]{\@secondoftwo}%
\providecommand \translation [1]{[#1]}%
\providecommand \BibitemOpen [0]{}%
\providecommand \bibitemStop [0]{}%
\providecommand \bibitemNoStop [0]{.\EOS\space}%
\providecommand \EOS [0]{\spacefactor3000\relax}%
\providecommand \BibitemShut  [1]{\csname bibitem#1\endcsname}%
\let\auto@bib@innerbib\@empty
%</preamble>
\bibitem [{\citenamefont {{Loeb}}\ and\ \citenamefont
  {{Waxman}}(2000)}]{LoebWaxman00}%
  \BibitemOpen
  \bibfield  {author} {\bibinfo {author} {\bibfnamefont {A.}~\bibnamefont
  {{Loeb}}}\ and\ \bibinfo {author} {\bibfnamefont {E.}~\bibnamefont
  {{Waxman}}},\ }\href@noop {} {\bibfield  {journal} {\bibinfo  {journal}
  {\nat}\ }\textbf {\bibinfo {volume} {405}},\ \bibinfo {pages} {156} (\bibinfo
  {year} {2000})},\ \Eprint {https://arxiv.org/abs/arXiv:astro-ph/0003447}
  {arXiv:astro-ph/0003447} \BibitemShut {NoStop}%
\bibitem [{\citenamefont {{Totani}}\ and\ \citenamefont
  {{Kitayama}}(2000)}]{TotaniKitayama00}%
  \BibitemOpen
  \bibfield  {author} {\bibinfo {author} {\bibfnamefont {T.}~\bibnamefont
  {{Totani}}}\ and\ \bibinfo {author} {\bibfnamefont {T.}~\bibnamefont
  {{Kitayama}}},\ }\href {https://doi.org/10.1086/317872} {\bibfield  {journal}
  {\bibinfo  {journal} {\apj}\ }\textbf {\bibinfo {volume} {545}},\ \bibinfo
  {pages} {572} (\bibinfo {year} {2000})},\ \Eprint
  {https://arxiv.org/abs/arXiv:astro-ph/0006176} {arXiv:astro-ph/0006176}
  \BibitemShut {NoStop}%
\bibitem [{\citenamefont {{Waxman}}\ and\ \citenamefont
  {{Loeb}}(2000)}]{WaxmanLoeb00}%
  \BibitemOpen
  \bibfield  {author} {\bibinfo {author} {\bibfnamefont {E.}~\bibnamefont
  {{Waxman}}}\ and\ \bibinfo {author} {\bibfnamefont {A.}~\bibnamefont
  {{Loeb}}},\ }\href {https://doi.org/10.1086/317326} {\bibfield  {journal}
  {\bibinfo  {journal} {\apjl}\ }\textbf {\bibinfo {volume} {545}},\ \bibinfo
  {pages} {L11} (\bibinfo {year} {2000})},\ \Eprint
  {https://arxiv.org/abs/arXiv:astro-ph/0007049} {arXiv:astro-ph/0007049}
  \BibitemShut {NoStop}%
\bibitem [{\citenamefont {{Keshet}}\ \emph {et~al.}(2003)\citenamefont
  {{Keshet}}, \citenamefont {{Waxman}}, \citenamefont {{Loeb}}, \citenamefont
  {{Springel}},\ and\ \citenamefont {{Hernquist}}}]{KeshetEtAl03}%
  \BibitemOpen
  \bibfield  {author} {\bibinfo {author} {\bibfnamefont {U.}~\bibnamefont
  {{Keshet}}}, \bibinfo {author} {\bibfnamefont {E.}~\bibnamefont {{Waxman}}},
  \bibinfo {author} {\bibfnamefont {A.}~\bibnamefont {{Loeb}}}, \emph
  {et~al.},\ }\href {https://doi.org/10.1086/345946} {\bibfield  {journal}
  {\bibinfo  {journal} {\apj}\ }\textbf {\bibinfo {volume} {585}},\ \bibinfo
  {pages} {128} (\bibinfo {year} {2003})},\ \Eprint
  {https://arxiv.org/abs/arXiv:astro-ph/0202318} {arXiv:astro-ph/0202318}
  \BibitemShut {NoStop}%
\bibitem [{\citenamefont {{Miniati}}(2002)}]{Miniati02}%
  \BibitemOpen
  \bibfield  {author} {\bibinfo {author} {\bibfnamefont {F.}~\bibnamefont
  {{Miniati}}},\ }\href {https://doi.org/10.1046/j.1365-8711.2002.05903.x}
  {\bibfield  {journal} {\bibinfo  {journal} {\mnras}\ }\textbf {\bibinfo
  {volume} {337}},\ \bibinfo {pages} {199} (\bibinfo {year} {2002})},\ \Eprint
  {https://arxiv.org/abs/arXiv:astro-ph/0203014} {arXiv:astro-ph/0203014}
  \BibitemShut {NoStop}%
\bibitem [{\citenamefont {{Keshet}}\ \emph
  {et~al.}(2004{\natexlab{a}})\citenamefont {{Keshet}}, \citenamefont
  {{Waxman}},\ and\ \citenamefont {{Loeb}}}]{KeshetEtAl04}%
  \BibitemOpen
  \bibfield  {author} {\bibinfo {author} {\bibfnamefont {U.}~\bibnamefont
  {{Keshet}}}, \bibinfo {author} {\bibfnamefont {E.}~\bibnamefont {{Waxman}}},\
  and\ \bibinfo {author} {\bibfnamefont {A.}~\bibnamefont {{Loeb}}},\ }\href
  {https://doi.org/10.1086/424837} {\bibfield  {journal} {\bibinfo  {journal}
  {\apj}\ }\textbf {\bibinfo {volume} {617}},\ \bibinfo {pages} {281} (\bibinfo
  {year} {2004}{\natexlab{a}})},\ \Eprint
  {https://arxiv.org/abs/arXiv:astro-ph/0402320} {arXiv:astro-ph/0402320}
  \BibitemShut {NoStop}%
\bibitem [{\citenamefont {{Keshet}}\ \emph
  {et~al.}(2004{\natexlab{b}})\citenamefont {{Keshet}}, \citenamefont
  {{Waxman}},\ and\ \citenamefont {{Loeb}}}]{KeshetEtAl04_SKA}%
  \BibitemOpen
  \bibfield  {author} {\bibinfo {author} {\bibfnamefont {U.}~\bibnamefont
  {{Keshet}}}, \bibinfo {author} {\bibfnamefont {E.}~\bibnamefont {{Waxman}}},\
  and\ \bibinfo {author} {\bibfnamefont {A.}~\bibnamefont {{Loeb}}},\ }\href
  {https://doi.org/10.1016/j.newar.2004.09.032} {\bibfield  {journal} {\bibinfo
   {journal} {\nar}\ }\textbf {\bibinfo {volume} {48}},\ \bibinfo {pages}
  {1119} (\bibinfo {year} {2004}{\natexlab{b}})},\ \Eprint
  {https://arxiv.org/abs/astro-ph/0407243} {astro-ph/0407243} \BibitemShut
  {NoStop}%
\bibitem [{\citenamefont {{Kocsis}}\ \emph {et~al.}(2005)\citenamefont
  {{Kocsis}}, \citenamefont {{Haiman}},\ and\ \citenamefont
  {{Frei}}}]{KocsisEtAl05}%
  \BibitemOpen
  \bibfield  {author} {\bibinfo {author} {\bibfnamefont {B.}~\bibnamefont
  {{Kocsis}}}, \bibinfo {author} {\bibfnamefont {Z.}~\bibnamefont {{Haiman}}},\
  and\ \bibinfo {author} {\bibfnamefont {Z.}~\bibnamefont {{Frei}}},\ }\href
  {https://doi.org/10.1086/427975} {\bibfield  {journal} {\bibinfo  {journal}
  {\apj}\ }\textbf {\bibinfo {volume} {623}},\ \bibinfo {pages} {632} (\bibinfo
  {year} {2005})},\ \Eprint {https://arxiv.org/abs/arXiv:astro-ph/0409430}
  {arXiv:astro-ph/0409430} \BibitemShut {NoStop}%
\bibitem [{\citenamefont {{Reiss}}\ \emph {et~al.}(2017)\citenamefont
  {{Reiss}}, \citenamefont {{Mushkin}},\ and\ \citenamefont
  {{Keshet}}}]{ReissEtAl17}%
  \BibitemOpen
  \bibfield  {author} {\bibinfo {author} {\bibfnamefont {I.}~\bibnamefont
  {{Reiss}}}, \bibinfo {author} {\bibfnamefont {J.}~\bibnamefont {{Mushkin}}},\
  and\ \bibinfo {author} {\bibfnamefont {U.}~\bibnamefont {{Keshet}}},\ }in\
  \href@noop {} {\emph {\bibinfo {booktitle} {Proceedings of the 7th
  International Fermi Symposium}}}\ (\bibinfo {year} {2017})\ p.\ \bibinfo
  {pages} {163},\ \Eprint {https://arxiv.org/abs/1712.06591} {arXiv:1712.06591
  [astro-ph.HE]} \BibitemShut {NoStop}%
\bibitem [{\citenamefont {{Reiss}}\ and\ \citenamefont
  {{Keshet}}(2018)}]{reiss2018detection}%
  \BibitemOpen
  \bibfield  {author} {\bibinfo {author} {\bibfnamefont {I.}~\bibnamefont
  {{Reiss}}}\ and\ \bibinfo {author} {\bibfnamefont {U.}~\bibnamefont
  {{Keshet}}},\ }\href {https://doi.org/10.1088/1475-7516/2018/10/010}
  {\bibfield  {journal} {\bibinfo  {journal} {\jcap}\ }\textbf {\bibinfo
  {volume} {2018}},\ \bibinfo {eid} {010} (\bibinfo {year} {2018})},\ \Eprint
  {https://arxiv.org/abs/1705.05376} {arXiv:1705.05376 [astro-ph.HE]}
  \BibitemShut {NoStop}%
\bibitem [{\citenamefont {{Hou}}\ \emph {et~al.}(2023)\citenamefont {{Hou}},
  \citenamefont {{Hallinan}},\ and\ \citenamefont {{Keshet}}}]{HouEtAl23}%
  \BibitemOpen
  \bibfield  {author} {\bibinfo {author} {\bibfnamefont {K.-C.}\ \bibnamefont
  {{Hou}}}, \bibinfo {author} {\bibfnamefont {G.}~\bibnamefont {{Hallinan}}},\
  and\ \bibinfo {author} {\bibfnamefont {U.}~\bibnamefont {{Keshet}}},\ }\href
  {https://doi.org/10.1093/mnras/stad785} {\bibfield  {journal} {\bibinfo
  {journal} {\mnras}\ }\textbf {\bibinfo {volume} {521}},\ \bibinfo {pages}
  {5786} (\bibinfo {year} {2023})},\ \Eprint {https://arxiv.org/abs/2210.09317}
  {arXiv:2210.09317 [astro-ph.HE]} \BibitemShut {NoStop}%
\bibitem [{\citenamefont {{Keshet}}(2025{\natexlab{a}})}]{Keshet25PaperII}%
  \BibitemOpen
  \bibfield  {author} {\bibinfo {author} {\bibfnamefont {U.}~\bibnamefont
  {{Keshet}}},\ }\href {https://doi.org/10.48550/arXiv.2502.19494} {\bibfield
  {journal} {\bibinfo  {journal} {arXiv e-prints}\ ,\ \bibinfo {eid}
  {arXiv:2502.19494}} (\bibinfo {year} {2025}{\natexlab{a}})},\ \Eprint
  {https://arxiv.org/abs/2502.19494} {arXiv:2502.19494 [astro-ph.HE]}
  \BibitemShut {NoStop}%
\bibitem [{\citenamefont {{Keshet}}(2025{\natexlab{b}})}]{Keshet25PaperIII}%
  \BibitemOpen
  \bibfield  {author} {\bibinfo {author} {\bibfnamefont {U.}~\bibnamefont
  {{Keshet}}},\ }\href@noop {} {\bibfield  {journal} {\bibinfo  {journal}
  {arXiv e-prints}\ ,\ \bibinfo {eid} {arXiv:2503.09687}} (\bibinfo {year}
  {2025}{\natexlab{b}})},\ \Eprint {https://arxiv.org/abs/2503.09687}
  {arXiv:2503.09687 [astro-ph.HE]} \BibitemShut {NoStop}%
\bibitem [{\citenamefont {{Huber}}\ \emph {et~al.}(2013)\citenamefont
  {{Huber}}, \citenamefont {{Tchernin}}, \citenamefont {{Eckert}},
  \citenamefont {{Farnier}}, \citenamefont {{Manalaysay}}, \citenamefont
  {{Straumann}},\ and\ \citenamefont {{Walter}}}]{HuberEtAl13}%
  \BibitemOpen
  \bibfield  {author} {\bibinfo {author} {\bibfnamefont {B.}~\bibnamefont
  {{Huber}}}, \bibinfo {author} {\bibfnamefont {C.}~\bibnamefont {{Tchernin}}},
  \bibinfo {author} {\bibfnamefont {D.}~\bibnamefont {{Eckert}}}, \emph
  {et~al.},\ }\href {https://doi.org/10.1051/0004-6361/201321947} {\bibfield
  {journal} {\bibinfo  {journal} {\aap}\ }\textbf {\bibinfo {volume} {560}},\
  \bibinfo {eid} {A64} (\bibinfo {year} {2013})},\ \Eprint
  {https://arxiv.org/abs/1308.6278} {arXiv:1308.6278 [astro-ph.HE]}
  \BibitemShut {NoStop}%
\bibitem [{\citenamefont {{Ackermann}}\ \emph {et~al.}(2014)\citenamefont
  {{Ackermann}}, \citenamefont {{Ajello}}, \citenamefont {{Albert}},
  \citenamefont {{Allafort}}, \citenamefont {{Atwood}}, \citenamefont
  {{Baldini}}, \citenamefont {{Ballet}}, \citenamefont {{Barbiellini}},
  \citenamefont {{Bastieri}}, \citenamefont {{Bechtol}}, \citenamefont
  {{Bellazzini}}, \citenamefont {{Bloom}}, \citenamefont {{Bonamente}},
  \citenamefont {{Bottacini}}, \citenamefont {{Brandt}}, \citenamefont
  {{Bregeon}}, \citenamefont {{Brigida}}, \citenamefont {{Bruel}},
  \citenamefont {{Buehler}}, \citenamefont {{Buson}}, \citenamefont
  {{Caliandro}}, \citenamefont {{Cameron}}, \citenamefont {{Caraveo}},
  \citenamefont {{Cavazzuti}}, \citenamefont {{Chaves}}, \citenamefont
  {{Chiang}}, \citenamefont {{Chiaro}}, \citenamefont {{Ciprini}},
  \citenamefont {{Claus}}, \citenamefont {{Cohen-Tanugi}}, \citenamefont
  {{Conrad}}, \citenamefont {{D'Ammando}}, \citenamefont {{de Angelis}},
  \citenamefont {{de Palma}}, \citenamefont {{Dermer}}, \citenamefont
  {{Digel}}, \citenamefont {{Drell}}, \citenamefont {{Drlica-Wagner}},
  \citenamefont {{Favuzzi}}, \citenamefont {{Franckowiak}}, \citenamefont
  {{Funk}}, \citenamefont {{Fusco}}, \citenamefont {{Gargano}}, \citenamefont
  {{Gasparrini}}, \citenamefont {{Germani}}, \citenamefont {{Giglietto}},
  \citenamefont {{Giordano}}, \citenamefont {{Giroletti}}, \citenamefont
  {{Godfrey}}, \citenamefont {{Gomez-Vargas}}, \citenamefont {{Grenier}},
  \citenamefont {{Guiriec}}, \citenamefont {{Gustafsson}}, \citenamefont
  {{Hadasch}}, \citenamefont {{Hayashida}}, \citenamefont {{Hewitt}},
  \citenamefont {{Hughes}}, \citenamefont {{Jeltema}}, \citenamefont
  {{J{\'o}hannesson}}, \citenamefont {{Johnson}}, \citenamefont {{Kamae}},
  \citenamefont {{Kataoka}}, \citenamefont {{Kn{\"o}dlseder}}, \citenamefont
  {{Kuss}}, \citenamefont {{Lande}}, \citenamefont {{Larsson}}, \citenamefont
  {{Latronico}}, \citenamefont {{Llena Garde}}, \citenamefont {{Longo}},
  \citenamefont {{Loparco}}, \citenamefont {{Lovellette}}, \citenamefont
  {{Lubrano}}, \citenamefont {{Mayer}}, \citenamefont {{Mazziotta}},
  \citenamefont {{McEnery}}, \citenamefont {{Michelson}}, \citenamefont
  {{Mitthumsiri}}, \citenamefont {{Mizuno}}, \citenamefont {{Monzani}},
  \citenamefont {{Morselli}}, \citenamefont {{Moskalenko}}, \citenamefont
  {{Murgia}}, \citenamefont {{Nemmen}}, \citenamefont {{Nuss}}, \citenamefont
  {{Ohsugi}}, \citenamefont {{Orienti}}, \citenamefont {{Orlando}},
  \citenamefont {{Ormes}}, \citenamefont {{Perkins}}, \citenamefont
  {{Pesce-Rollins}}, \citenamefont {{Piron}}, \citenamefont {{Pivato}},
  \citenamefont {{Rain{\`o}}}, \citenamefont {{Rando}}, \citenamefont
  {{Razzano}}, \citenamefont {{Razzaque}}, \citenamefont {{Reimer}},
  \citenamefont {{Reimer}}, \citenamefont {{Ruan}}, \citenamefont
  {{S{\'a}nchez-Conde}}, \citenamefont {{Schulz}}, \citenamefont {{Sgr{\`o}}},
  \citenamefont {{Siskind}}, \citenamefont {{Spandre}}, \citenamefont
  {{Spinelli}}, \citenamefont {{Storm}}, \citenamefont {{Strong}},
  \citenamefont {{Suson}}, \citenamefont {{Takahashi}}, \citenamefont
  {{Thayer}}, \citenamefont {{Thayer}}, \citenamefont {{Thompson}},
  \citenamefont {{Tibaldo}}, \citenamefont {{Tinivella}}, \citenamefont
  {{Torres}}, \citenamefont {{Troja}}, \citenamefont {{Uchiyama}},
  \citenamefont {{Usher}}, \citenamefont {{Vandenbroucke}}, \citenamefont
  {{Vianello}}, \citenamefont {{Vitale}}, \citenamefont {{Winer}},
  \citenamefont {{Wood}}, \citenamefont {{Zimmer}}, \citenamefont {{Fermi-LAT
  Collaboration}}, \citenamefont {{Pinzke}},\ and\ \citenamefont
  {{Pfrommer}}}]{AckermannEtAl14_GammaRayLimits}%
  \BibitemOpen
  \bibfield  {author} {\bibinfo {author} {\bibfnamefont {M.}~\bibnamefont
  {{Ackermann}}}, \bibinfo {author} {\bibfnamefont {M.}~\bibnamefont
  {{Ajello}}}, \bibinfo {author} {\bibfnamefont {A.}~\bibnamefont {{Albert}}},
  \emph {et~al.},\ }\href {https://doi.org/10.1088/0004-637X/787/1/18}
  {\bibfield  {journal} {\bibinfo  {journal} {\apj}\ }\textbf {\bibinfo
  {volume} {787}},\ \bibinfo {eid} {18} (\bibinfo {year} {2014})},\ \Eprint
  {https://arxiv.org/abs/1308.5654} {arXiv:1308.5654 [astro-ph.HE]}
  \BibitemShut {NoStop}%
\bibitem [{\citenamefont {{Prokhorov}}\ and\ \citenamefont
  {{Churazov}}(2014)}]{ProkhorovChurazov14}%
  \BibitemOpen
  \bibfield  {author} {\bibinfo {author} {\bibfnamefont {D.~A.}\ \bibnamefont
  {{Prokhorov}}}\ and\ \bibinfo {author} {\bibfnamefont {E.~M.}\ \bibnamefont
  {{Churazov}}},\ }\href {https://doi.org/10.1051/0004-6361/201322454}
  {\bibfield  {journal} {\bibinfo  {journal} {\aap}\ }\textbf {\bibinfo
  {volume} {567}},\ \bibinfo {eid} {A93} (\bibinfo {year} {2014})},\ \Eprint
  {https://arxiv.org/abs/1309.0197} {arXiv:1309.0197 [astro-ph.HE]}
  \BibitemShut {NoStop}%
\bibitem [{\citenamefont {{Griffin}}\ \emph {et~al.}(2014)\citenamefont
  {{Griffin}}, \citenamefont {{Dai}},\ and\ \citenamefont
  {{Kochanek}}}]{GriffinEtAl14}%
  \BibitemOpen
  \bibfield  {author} {\bibinfo {author} {\bibfnamefont {R.~D.}\ \bibnamefont
  {{Griffin}}}, \bibinfo {author} {\bibfnamefont {X.}~\bibnamefont {{Dai}}},\
  and\ \bibinfo {author} {\bibfnamefont {C.~S.}\ \bibnamefont {{Kochanek}}},\
  }\href@noop {} {\bibfield  {journal} {\bibinfo  {journal} {ArXiv e-prints}\ }
  (\bibinfo {year} {2014})},\ \Eprint {https://arxiv.org/abs/1405.7047}
  {arXiv:1405.7047 [astro-ph.HE]} \BibitemShut {NoStop}%
\bibitem [{\citenamefont {{Piffaretti}}\ \emph {et~al.}(2011)\citenamefont
  {{Piffaretti}}, \citenamefont {{Arnaud}}, \citenamefont {{Pratt}},
  \citenamefont {{Pointecouteau}},\ and\ \citenamefont
  {{Melin}}}]{PiffarettiEtAl11}%
  \BibitemOpen
  \bibfield  {author} {\bibinfo {author} {\bibfnamefont {R.}~\bibnamefont
  {{Piffaretti}}}, \bibinfo {author} {\bibfnamefont {M.}~\bibnamefont
  {{Arnaud}}}, \bibinfo {author} {\bibfnamefont {G.~W.}\ \bibnamefont
  {{Pratt}}}, \emph {et~al.},\ }\href
  {https://doi.org/10.1051/0004-6361/201015377} {\bibfield  {journal} {\bibinfo
   {journal} {\aap}\ }\textbf {\bibinfo {volume} {534}},\ \bibinfo {eid} {A109}
  (\bibinfo {year} {2011})},\ \Eprint {https://arxiv.org/abs/1007.1916}
  {arXiv:1007.1916} \BibitemShut {NoStop}%
\bibitem [{\citenamefont {{Ilani}}\ \emph {et~al.}(2024)\citenamefont
  {{Ilani}}, \citenamefont {{Hou}}, \citenamefont {{Nadler}},\ and\
  \citenamefont {{Keshet}}}]{IlaniEtAl24}%
  \BibitemOpen
  \bibfield  {author} {\bibinfo {author} {\bibfnamefont {G.}~\bibnamefont
  {{Ilani}}}, \bibinfo {author} {\bibfnamefont {K.-C.}\ \bibnamefont {{Hou}}},
  \bibinfo {author} {\bibfnamefont {G.}~\bibnamefont {{Nadler}}},\ and\
  \bibinfo {author} {\bibfnamefont {U.}~\bibnamefont {{Keshet}}},\ }\href
  {https://doi.org/10.1051/0004-6361/202449819} {\bibfield  {journal} {\bibinfo
   {journal} {\aap}\ }\textbf {\bibinfo {volume} {686}},\ \bibinfo {eid} {L16}
  (\bibinfo {year} {2024})},\ \Eprint {https://arxiv.org/abs/2402.17822}
  {arXiv:2402.17822 [astro-ph.HE]} \BibitemShut {NoStop}%
\bibitem [{\citenamefont {{Nadler}}\ \emph {et~al.}(2024)\citenamefont
  {{Nadler}}, \citenamefont {{Hou}},\ and\ \citenamefont
  {{Keshet}}}]{Nadler24InPrep}%
  \BibitemOpen
  \bibfield  {author} {\bibinfo {author} {\bibfnamefont {G.}~\bibnamefont
  {{Nadler}}}, \bibinfo {author} {\bibfnamefont {K.-C.}\ \bibnamefont
  {{Hou}}},\ and\ \bibinfo {author} {\bibfnamefont {U.}~\bibnamefont
  {{Keshet}}}} (\bibinfo {year} {2024}),\ \bibinfo {note} {in
  preparation}\BibitemShut {NoStop}%
\bibitem [{\citenamefont {{Keshet}}\ \emph {et~al.}(2017)\citenamefont
  {{Keshet}}, \citenamefont {{Kushnir}}, \citenamefont {{Loeb}},\ and\
  \citenamefont {{Waxman}}}]{KeshetEtAl17}%
  \BibitemOpen
  \bibfield  {author} {\bibinfo {author} {\bibfnamefont {U.}~\bibnamefont
  {{Keshet}}}, \bibinfo {author} {\bibfnamefont {D.}~\bibnamefont {{Kushnir}}},
  \bibinfo {author} {\bibfnamefont {A.}~\bibnamefont {{Loeb}}},\ and\ \bibinfo
  {author} {\bibfnamefont {E.}~\bibnamefont {{Waxman}}},\ }\href
  {https://doi.org/10.3847/1538-4357/aa794b} {\bibfield  {journal} {\bibinfo
  {journal} {\apj}\ }\textbf {\bibinfo {volume} {845}},\ \bibinfo {eid} {24}
  (\bibinfo {year} {2017})}\BibitemShut {NoStop}%
\bibitem [{\citenamefont {{Keshet}}\ and\ \citenamefont
  {{Reiss}}(2018)}]{keshet2018evidence}%
  \BibitemOpen
  \bibfield  {author} {\bibinfo {author} {\bibfnamefont {U.}~\bibnamefont
  {{Keshet}}}\ and\ \bibinfo {author} {\bibfnamefont {I.}~\bibnamefont
  {{Reiss}}},\ }\href {https://doi.org/10.3847/1538-4357/aaeb1d} {\bibfield
  {journal} {\bibinfo  {journal} {\apj}\ }\textbf {\bibinfo {volume} {869}},\
  \bibinfo {eid} {53} (\bibinfo {year} {2018})}\BibitemShut {NoStop}%
\bibitem [{\citenamefont {{Hurier}}\ \emph {et~al.}(2019)\citenamefont
  {{Hurier}}, \citenamefont {{Adam}},\ and\ \citenamefont
  {{Keshet}}}]{HurierEtAl19}%
  \BibitemOpen
  \bibfield  {author} {\bibinfo {author} {\bibfnamefont {G.}~\bibnamefont
  {{Hurier}}}, \bibinfo {author} {\bibfnamefont {R.}~\bibnamefont {{Adam}}},\
  and\ \bibinfo {author} {\bibfnamefont {U.}~\bibnamefont {{Keshet}}},\ }\href
  {https://doi.org/10.1051/0004-6361/201732468} {\bibfield  {journal} {\bibinfo
   {journal} {\aap}\ }\textbf {\bibinfo {volume} {622}},\ \bibinfo {eid} {A136}
  (\bibinfo {year} {2019})}\BibitemShut {NoStop}%
\bibitem [{\citenamefont {{Keshet}}\ \emph {et~al.}(2020)\citenamefont
  {{Keshet}}, \citenamefont {{Reiss}},\ and\ \citenamefont
  {{Hurier}}}]{keshet20coincident}%
  \BibitemOpen
  \bibfield  {author} {\bibinfo {author} {\bibfnamefont {U.}~\bibnamefont
  {{Keshet}}}, \bibinfo {author} {\bibfnamefont {I.}~\bibnamefont {{Reiss}}},\
  and\ \bibinfo {author} {\bibfnamefont {G.}~\bibnamefont {{Hurier}}},\ }\href
  {https://doi.org/10.3847/1538-4357/ab8c49} {\bibfield  {journal} {\bibinfo
  {journal} {\apj}\ }\textbf {\bibinfo {volume} {895}},\ \bibinfo {eid} {72}
  (\bibinfo {year} {2020})},\ \Eprint {https://arxiv.org/abs/1801.01494}
  {arXiv:1801.01494 [astro-ph.CO]} \BibitemShut {NoStop}%
\bibitem [{\citenamefont {Ilani}\ \emph {et~al.}(2024)\citenamefont {Ilani},
  \citenamefont {Hou},\ and\ \citenamefont {Keshet}}]{IlaniEtAl24a}%
  \BibitemOpen
  \bibfield  {author} {\bibinfo {author} {\bibfnamefont {G.}~\bibnamefont
  {Ilani}}, \bibinfo {author} {\bibfnamefont {K.-C.}\ \bibnamefont {Hou}},\
  and\ \bibinfo {author} {\bibfnamefont {U.}~\bibnamefont {Keshet}},\ }\href
  {https://doi.org/10.1088/1475-7516/2024/10/008} {\bibfield  {journal}
  {\bibinfo  {journal} {Journal of Cosmology and Astroparticle Physics}\
  }\textbf {\bibinfo {volume} {2024}}\bibinfo  {number} { (10)},\ \bibinfo
  {pages} {008}}\BibitemShut {NoStop}%
\bibitem [{\citenamefont {{More}}\ \emph {et~al.}(2016)\citenamefont {{More}},
  \citenamefont {{Miyatake}}, \citenamefont {{Takada}}, \citenamefont
  {{Diemer}}, \citenamefont {{Kravtsov}}, \citenamefont {{Dalal}},
  \citenamefont {{More}}, \citenamefont {{Murata}}, \citenamefont
  {{Mandelbaum}}, \citenamefont {{Rozo}}, \citenamefont {{Rykoff}},
  \citenamefont {{Oguri}},\ and\ \citenamefont {{Spergel}}}]{MoreEtAl16}%
  \BibitemOpen
\bibfield  {number} {  }\bibfield  {author} {\bibinfo {author} {\bibfnamefont
  {S.}~\bibnamefont {{More}}}, \bibinfo {author} {\bibfnamefont
  {H.}~\bibnamefont {{Miyatake}}}, \bibinfo {author} {\bibfnamefont
  {M.}~\bibnamefont {{Takada}}}, \emph {et~al.},\ }\href
  {https://doi.org/10.3847/0004-637X/825/1/39} {\bibfield  {journal} {\bibinfo
  {journal} {\apj}\ }\textbf {\bibinfo {volume} {825}},\ \bibinfo {eid} {39}
  (\bibinfo {year} {2016})},\ \Eprint {https://arxiv.org/abs/1601.06063}
  {arXiv:1601.06063 [astro-ph.CO]} \BibitemShut {NoStop}%
\bibitem [{\citenamefont {{Shin}}\ \emph {et~al.}(2019)\citenamefont {{Shin}},
  \citenamefont {{Adhikari}}, \citenamefont {{Baxter}}, \citenamefont
  {{Chang}}, \citenamefont {{Jain}}, \citenamefont {{Battaglia}}, \citenamefont
  {{Bleem}}, \citenamefont {{Bocquet}}, \citenamefont {{DeRose}}, \citenamefont
  {{Gruen}}, \citenamefont {{Hilton}}, \citenamefont {{Kravtsov}},
  \citenamefont {{McClintock}}, \citenamefont {{Rozo}}, \citenamefont
  {{Rykoff}}, \citenamefont {{Varga}}, \citenamefont {{Wechsler}},
  \citenamefont {{Wu}}, \citenamefont {{Zhang}}, \citenamefont {{Aiola}},
  \citenamefont {{Allam}}, \citenamefont {{Bechtol}}, \citenamefont {{Benson}},
  \citenamefont {{Bertin}}, \citenamefont {{Bond}}, \citenamefont {{Brodwin}},
  \citenamefont {{Brooks}}, \citenamefont {{Buckley-Geer}}, \citenamefont
  {{Burke}}, \citenamefont {{Carlstrom}}, \citenamefont {{Carnero Rosell}},
  \citenamefont {{Carrasco Kind}}, \citenamefont {{Carretero}}, \citenamefont
  {{Castander}}, \citenamefont {{Choi}}, \citenamefont {{Cunha}}, \citenamefont
  {{Crawford}}, \citenamefont {{da Costa}}, \citenamefont {{De Vicente}},
  \citenamefont {{Desai}}, \citenamefont {{Devlin}}, \citenamefont
  {{Dietrich}}, \citenamefont {{Doel}}, \citenamefont {{Dunkley}},
  \citenamefont {{Eifler}}, \citenamefont {{Evrard}}, \citenamefont
  {{Flaugher}}, \citenamefont {{Fosalba}}, \citenamefont {{Gallardo}},
  \citenamefont {{Garc{\'\i}a-Bellido}}, \citenamefont {{Gaztanaga}},
  \citenamefont {{Gerdes}}, \citenamefont {{Gralla}}, \citenamefont
  {{Gruendl}}, \citenamefont {{Gschwend}}, \citenamefont {{Gupta}},
  \citenamefont {{Gutierrez}}, \citenamefont {{Hartley}}, \citenamefont
  {{Hill}}, \citenamefont {{Ho}}, \citenamefont {{Hollowood}}, \citenamefont
  {{Honscheid}}, \citenamefont {{Hoyle}}, \citenamefont {{Huffenberger}},
  \citenamefont {{Hughes}}, \citenamefont {{James}}, \citenamefont {{Jeltema}},
  \citenamefont {{Kim}}, \citenamefont {{Krause}}, \citenamefont {{Kuehn}},
  \citenamefont {{Lahav}}, \citenamefont {{Lima}}, \citenamefont
  {{Madhavacheril}}, \citenamefont {{Maia}}, \citenamefont {{Marshall}},
  \citenamefont {{Maurin}}, \citenamefont {{McMahon}}, \citenamefont
  {{Menanteau}}, \citenamefont {{Miller}}, \citenamefont {{Miquel}},
  \citenamefont {{Mohr}}, \citenamefont {{Naess}}, \citenamefont {{Nati}},
  \citenamefont {{Newburgh}}, \citenamefont {{Niemack}}, \citenamefont
  {{Ogando}}, \citenamefont {{Page}}, \citenamefont {{Partridge}},
  \citenamefont {{Patil}}, \citenamefont {{Plazas}}, \citenamefont {{Rapetti}},
  \citenamefont {{Reichardt}}, \citenamefont {{Romer}}, \citenamefont
  {{Sanchez}}, \citenamefont {{Scarpine}}, \citenamefont {{Schindler}},
  \citenamefont {{Serrano}}, \citenamefont {{Smith}}, \citenamefont {{Smith}},
  \citenamefont {{Soares-Santos}}, \citenamefont {{Sobreira}}, \citenamefont
  {{Staggs}}, \citenamefont {{Stark}}, \citenamefont {{Stein}}, \citenamefont
  {{Suchyta}}, \citenamefont {{Swanson}}, \citenamefont {{Tarle}},
  \citenamefont {{Thomas}}, \citenamefont {{van Engelen}}, \citenamefont
  {{Wollack}},\ and\ \citenamefont {{Xu}}}]{ShinEtAl19}%
  \BibitemOpen
  \bibfield  {author} {\bibinfo {author} {\bibfnamefont {T.}~\bibnamefont
  {{Shin}}}, \bibinfo {author} {\bibfnamefont {S.}~\bibnamefont {{Adhikari}}},
  \bibinfo {author} {\bibfnamefont {E.~J.}\ \bibnamefont {{Baxter}}}, \emph
  {et~al.},\ }\href {https://doi.org/10.1093/mnras/stz1434} {\bibfield
  {journal} {\bibinfo  {journal} {\mnras}\ }\textbf {\bibinfo {volume} {487}},\
  \bibinfo {pages} {2900} (\bibinfo {year} {2019})},\ \Eprint
  {https://arxiv.org/abs/1811.06081} {arXiv:1811.06081 [astro-ph.CO]}
  \BibitemShut {NoStop}%
\bibitem [{\citenamefont {{Wolleben}}\ \emph {et~al.}(2021)\citenamefont
  {{Wolleben}}, \citenamefont {{Landecker}}, \citenamefont {{Douglas}},
  \citenamefont {{Gray}}, \citenamefont {{Ordog}}, \citenamefont {{Dickey}},
  \citenamefont {{Hill}}, \citenamefont {{Carretti}}, \citenamefont {{Brown}},
  \citenamefont {{Gaensler}}, \citenamefont {{Han}}, \citenamefont
  {{Haverkorn}}, \citenamefont {{Kothes}}, \citenamefont {{Leahy}},
  \citenamefont {{McClure-Griffiths}}, \citenamefont {{McConnell}},
  \citenamefont {{Reich}}, \citenamefont {{Taylor}}, \citenamefont
  {{Thomson}},\ and\ \citenamefont {{West}}}]{WollebenEtAl21}%
  \BibitemOpen
  \bibfield  {author} {\bibinfo {author} {\bibfnamefont {M.}~\bibnamefont
  {{Wolleben}}}, \bibinfo {author} {\bibfnamefont {T.~L.}\ \bibnamefont
  {{Landecker}}}, \bibinfo {author} {\bibfnamefont {K.~A.}\ \bibnamefont
  {{Douglas}}}, \emph {et~al.},\ }\href
  {https://doi.org/10.3847/1538-3881/abf7c1} {\bibfield  {journal} {\bibinfo
  {journal} {\aj}\ }\textbf {\bibinfo {volume} {162}},\ \bibinfo {eid} {35}
  (\bibinfo {year} {2021})},\ \Eprint {https://arxiv.org/abs/2106.00945}
  {arXiv:2106.00945 [astro-ph.GA]} \BibitemShut {NoStop}%
\bibitem [{\citenamefont {{Taylor}}\ \emph {et~al.}(2009)\citenamefont
  {{Taylor}}, \citenamefont {{Stil}},\ and\ \citenamefont
  {{Sunstrum}}}]{TaylorEtAl09}%
  \BibitemOpen
  \bibfield  {author} {\bibinfo {author} {\bibfnamefont {A.~R.}\ \bibnamefont
  {{Taylor}}}, \bibinfo {author} {\bibfnamefont {J.~M.}\ \bibnamefont
  {{Stil}}},\ and\ \bibinfo {author} {\bibfnamefont {C.}~\bibnamefont
  {{Sunstrum}}},\ }\href {https://doi.org/10.1088/0004-637X/702/2/1230}
  {\bibfield  {journal} {\bibinfo  {journal} {\apj}\ }\textbf {\bibinfo
  {volume} {702}},\ \bibinfo {pages} {1230} (\bibinfo {year}
  {2009})}\BibitemShut {NoStop}%
\bibitem [{Note1()}]{Note1}%
  \BibitemOpen
  \bibinfo {note} {With polarization angle in IAU conventions (T. Landecker,
  private communications): counter-clockwise on the plane of the sky, north due
  east.}\BibitemShut {Stop}%
\bibitem [{Note2()}]{Note2}%
  \BibitemOpen
  \bibinfo {note} {The angle $\phi $ defined with polarization-angle
  conventions: north due east.}\BibitemShut {Stop}%
\bibitem [{\citenamefont {{Condon}}\ \emph {et~al.}(1998)\citenamefont
  {{Condon}}, \citenamefont {{Cotton}}, \citenamefont {{Greisen}},
  \citenamefont {{Yin}}, \citenamefont {{Perley}}, \citenamefont {{Taylor}},\
  and\ \citenamefont {{Broderick}}}]{NVSS_paper}%
  \BibitemOpen
  \bibfield  {author} {\bibinfo {author} {\bibfnamefont {J.~J.}\ \bibnamefont
  {{Condon}}}, \bibinfo {author} {\bibfnamefont {W.~D.}\ \bibnamefont
  {{Cotton}}}, \bibinfo {author} {\bibfnamefont {E.~W.}\ \bibnamefont
  {{Greisen}}}, \emph {et~al.},\ }\href {https://doi.org/10.1086/300337}
  {\bibfield  {journal} {\bibinfo  {journal} {\aj}\ }\textbf {\bibinfo {volume}
  {115}},\ \bibinfo {pages} {1693} (\bibinfo {year} {1998})}\BibitemShut
  {NoStop}%
\bibitem [{\citenamefont {{Duchesne}}\ \emph {et~al.}(2024)\citenamefont
  {{Duchesne}}, \citenamefont {{Grundy}}, \citenamefont {{Heald}},
  \citenamefont {{Lenc}}, \citenamefont {{Leung}}, \citenamefont {{McConnell}},
  \citenamefont {{Murphy}}, \citenamefont {{Pritchard}}, \citenamefont
  {{Rose}}, \citenamefont {{Thomson}}, \citenamefont {{Wang}}, \citenamefont
  {{Wang}},\ and\ \citenamefont {{Whiting}}}]{DuchesneEtAl24}%
  \BibitemOpen
  \bibfield  {author} {\bibinfo {author} {\bibfnamefont {S.~W.}\ \bibnamefont
  {{Duchesne}}}, \bibinfo {author} {\bibfnamefont {J.~A.}\ \bibnamefont
  {{Grundy}}}, \bibinfo {author} {\bibfnamefont {G.~H.}\ \bibnamefont
  {{Heald}}}, \emph {et~al.},\ }\href {https://doi.org/10.1017/pasa.2023.60}
  {\bibfield  {journal} {\bibinfo  {journal} {\pasa}\ }\textbf {\bibinfo
  {volume} {41}},\ \bibinfo {eid} {e003} (\bibinfo {year} {2024})},\ \Eprint
  {https://arxiv.org/abs/2311.12369} {arXiv:2311.12369 [astro-ph.GA]}
  \BibitemShut {NoStop}%
\bibitem [{\citenamefont {Wilks}(1938)}]{Wilks1938}%
  \BibitemOpen
  \bibfield  {author} {\bibinfo {author} {\bibfnamefont {S.~S.}\ \bibnamefont
  {Wilks}},\ }\href {http://www.jstor.org/stable/2957648} {\bibfield  {journal}
  {\bibinfo  {journal} {The Annals of Mathematical Statistics}\ }\textbf
  {\bibinfo {volume} {9}},\ \bibinfo {pages} {60} (\bibinfo {year}
  {1938})}\BibitemShut {NoStop}%
\bibitem [{\citenamefont {{Keshet}}\ and\ \citenamefont
  {{Hou}}(2024)}]{KeshetHou24}%
  \BibitemOpen
  \bibfield  {author} {\bibinfo {author} {\bibfnamefont {U.}~\bibnamefont
  {{Keshet}}}\ and\ \bibinfo {author} {\bibfnamefont {K.-C.}\ \bibnamefont
  {{Hou}}},\ }\href {https://doi.org/10.48550/arXiv.2412.03645} {\bibfield
  {journal} {\bibinfo  {journal} {arXiv e-prints}\ ,\ \bibinfo {eid}
  {arXiv:2412.03645}} (\bibinfo {year} {2024})},\ \Eprint
  {https://arxiv.org/abs/2412.03645} {arXiv:2412.03645 [astro-ph.HE]}
  \BibitemShut {NoStop}%
\bibitem [{\citenamefont {{Vernstrom}}\ \emph {et~al.}(2023)\citenamefont
  {{Vernstrom}}, \citenamefont {{West}}, \citenamefont {{Vazza}}, \citenamefont
  {{Wittor}}, \citenamefont {{Riseley}},\ and\ \citenamefont
  {{Heald}}}]{VernstromEtAl23}%
  \BibitemOpen
  \bibfield  {author} {\bibinfo {author} {\bibfnamefont {T.}~\bibnamefont
  {{Vernstrom}}}, \bibinfo {author} {\bibfnamefont {J.}~\bibnamefont {{West}}},
  \bibinfo {author} {\bibfnamefont {F.}~\bibnamefont {{Vazza}}}, \emph
  {et~al.},\ }\href {https://doi.org/10.1126/sciadv.ade7233} {\bibfield
  {journal} {\bibinfo  {journal} {Science Advances}\ }\textbf {\bibinfo
  {volume} {9}},\ \bibinfo {eid} {eade7233} (\bibinfo {year} {2023})},\ \Eprint
  {https://arxiv.org/abs/2302.08072} {arXiv:2302.08072 [astro-ph.CO]}
  \BibitemShut {NoStop}%
\bibitem [{\citenamefont {{Keshet}}(2010)}]{Keshet10}%
  \BibitemOpen
  \bibfield  {author} {\bibinfo {author} {\bibfnamefont {U.}~\bibnamefont
  {{Keshet}}},\ }\href@noop {} {\bibfield  {journal} {\bibinfo  {journal}
  {ArXiv e-prints}\ } (\bibinfo {year} {2010})},\ \Eprint
  {https://arxiv.org/abs/1011.0729} {arXiv:1011.0729 [astro-ph.HE]}
  \BibitemShut {NoStop}%
\bibitem [{\citenamefont {{Keshet}}(2024)}]{Keshet24}%
  \BibitemOpen
  \bibfield  {author} {\bibinfo {author} {\bibfnamefont {U.}~\bibnamefont
  {{Keshet}}},\ }\href {https://doi.org/10.1093/mnras/stad3154} {\bibfield
  {journal} {\bibinfo  {journal} {\mnras}\ }\textbf {\bibinfo {volume} {527}},\
  \bibinfo {pages} {1194} (\bibinfo {year} {2024})},\ \Eprint
  {https://arxiv.org/abs/2303.08146} {arXiv:2303.08146 [astro-ph.HE]}
  \BibitemShut {NoStop}%
\bibitem [{\citenamefont {{Keshet}}(2025{\natexlab{c}})}]{Keshet25Phoenix}%
  \BibitemOpen
  \bibfield  {author} {\bibinfo {author} {\bibfnamefont {U.}~\bibnamefont
  {{Keshet}}},\ }\href@noop {} {\bibfield  {journal} {\bibinfo  {journal}
  {arXiv e-prints}\ ,\ \bibinfo {eid} {arXiv:2503.07714}} (\bibinfo {year}
  {2025}{\natexlab{c}})},\ \Eprint {https://arxiv.org/abs/2503.07714}
  {arXiv:2503.07714 [astro-ph.HE]} \BibitemShut {NoStop}%
\bibitem [{\citenamefont {{van Weeren}}\ \emph {et~al.}(2019)\citenamefont
  {{van Weeren}}, \citenamefont {{de Gasperin}}, \citenamefont {{Akamatsu}},
  \citenamefont {{Br{\"u}ggen}}, \citenamefont {{Feretti}}, \citenamefont
  {{Kang}}, \citenamefont {{Stroe}},\ and\ \citenamefont
  {{Zandanel}}}]{vanWeerenEtAl19}%
  \BibitemOpen
  \bibfield  {author} {\bibinfo {author} {\bibfnamefont {R.~J.}\ \bibnamefont
  {{van Weeren}}}, \bibinfo {author} {\bibfnamefont {F.}~\bibnamefont {{de
  Gasperin}}}, \bibinfo {author} {\bibfnamefont {H.}~\bibnamefont
  {{Akamatsu}}}, \emph {et~al.},\ }\href
  {https://doi.org/10.1007/s11214-019-0584-z} {\bibfield  {journal} {\bibinfo
  {journal} {\ssr}\ }\textbf {\bibinfo {volume} {215}},\ \bibinfo {eid} {16}
  (\bibinfo {year} {2019})},\ \Eprint {https://arxiv.org/abs/1901.04496}
  {arXiv:1901.04496 [astro-ph.HE]} \BibitemShut {NoStop}%
\end{thebibliography}%

\appendix

\begin{widetext}
\section{Modelling the stacked shock signature}
\label{sec:Projection}

Consider a radiation source with an arbitrary 3D emissivity distribution $j_\nu(\bm{r})$.
Projecting and averaging over all possible source orientations yields the mean, stacked, radial brightness profile
\begin{equation}\label{eq:stackedI}
  I_\nu(b) = \frac{1}{2\pi}\int \frac{r \,\Theta(r-b)}{\sqrt{r^2-b^2}} j_\nu(\bm{r}) \,\sin(\theta) \,d\varphi \,d\theta \,dr \coma
\end{equation}
as a function of impact parameter $b=R_{500}\tau$.
Here, $\Theta$ is the Heaviside step function, we use spherical coordinates $(r,\theta,\varphi)$ without any assumptions on $j_\nu(\bm{r})$, and redshift effects are neglected.
If the source is axially symmetric, azimuthal integration over $\varphi$ cancels the factor $2\pi$ once the symmetry axis is chosen along polar angle $\theta=0$.
For simple axisymmetric $j_\nu(\bm{r})$ distributions, such as prolate or oblate spheroids, integration over $\theta$ can then be carried out analytically; in some cases, radial integration can be performed analytically as well.

Consider for instance a cylindrical accretion shock of half-height $h$ and small base radius $\zeta h<h$, radiating within a layer of thickness $\delta$ downstream. The integral \eqref{eq:stackedI} can be carried out analytically in the limit of an infinitely thin emission layer, $\delta\to 0$, if the emissivity integrated across the layer is approximately uniform, $j_\nu\,\delta\simeq \const$. The curved cylinder wall then contributes
\begin{equation}\label{eq:Cylinder1}
  \frac{I_\nu(b)}{j_\nu\,\delta} = \frac{\zeta}{\pi h}\Re\left( F\left\{\arcsin\left[\max\left(1,\frac{\zeta}{b}\right)\right]\,\middle|\, \frac{b^2}{\zeta^2}\right\}
  -F\left\{\arccsc\left[\frac{b}{\sqrt{1+\zeta^2}}\right]\,\middle|\,\frac{b^2}{\zeta^2}\right\}\right)\coma
\end{equation}
where $F(\phi|m)$ is the elliptic integral of the first kind, whereas the bases of the cylinder jointly contribute
\begin{equation}\label{eq:Cylinder2}
\frac{I_\nu(b)}{j_\nu\,\delta} =
\frac{1}{2\pi}\times
\begin{cases}
 \log \left(\sqrt{\frac{1+\zeta^2}{1+\zeta^2-b^2}}+1\right) - \log \left(\sqrt{\frac{1+\zeta^2}{1+\zeta^2-b^2}}-1\right) & \mbox{ if }b>1 \, ;\\
 \log \left[\frac{\left(1-\sqrt{1-b^2}\right) \left(\sqrt{1+\zeta^2-b^2}+\sqrt{1+\zeta^2}\right)^2}{\left(1+\sqrt{1-b^2}\right)b^2}\right] & \mbox{ if }b<1 \, .
\end{cases}
\end{equation}
Generalization is straightforward for different values of $j_\nu\,\delta$ at each base and along the curved wall, for a wide, $\zeta>1$ base, etc.
\end{widetext}

\end{document}